\documentclass[superscriptaddress,aps,prx,reprint,nofootinbib,notitlepage,10pt,longbibliography]{revtex4-2}
\pdfoutput=1
\usepackage{graphicx}

\usepackage{multirow}
\usepackage{subcaption}
\usepackage{amsmath}
\usepackage{adjustbox}
\usepackage{amsfonts}
\usepackage{amsthm}
\usepackage{amssymb}
\usepackage[justification=justified,singlelinecheck=false,compatibility=false]{caption}
\usepackage{makecell,booktabs}

\usepackage{tabularx}
\usepackage{mathtools}
\usepackage{siunitx}  

\usepackage[table]{xcolor}

\usepackage[edges]{forest}
\usetikzlibrary{shapes.geometric, arrows}

\usepackage[colorlinks]{hyperref}
\hypersetup{
	pdfstartview={FitH},
	pdfnewwindow=true,
	colorlinks=true,
	linkcolor=blue,
	citecolor=blue,
	filecolor=blue,
	urlcolor=blue}
\usepackage[all]{hypcap}

\usepackage{textcomp}
\usepackage{float}

\usepackage{ragged2e}
\makeatletter
\long\def\@makecaption#1#2{%
  \vskip\abovecaptionskip
  \sbox\@tempboxa{#1: #2}%
  \ifdim \wd\@tempboxa >\hsize
    \justifying\normalfont\small \noindent #1: #2\par
  \else
    \global \@minipagefalse
    \hb@xt@\hsize{\hfil\box\@tempboxa\hfil}%
  \fi}
\makeatother

\usepackage[utf8]{inputenc}

\definecolor{codegreen}{rgb}{0,0.6,0}
\definecolor{codegray}{rgb}{0.5,0.5,0.5}
\definecolor{codepurple}{rgb}{0.58,0,0.82}
\definecolor{backcolour}{rgb}{0.95,0.95,0.92}

\usepackage{colortbl}

\renewcommand{\arraystretch}{1.1}

\def\ket#1{\mathinner{|{#1}\rangle}}

\newlength{\oldwidth}
\newlength{\oldheight}
\begin{document}
\makeatletter
\def\l@subsubsection#1#2{}
\def\l@f@section{}
\def\toc@@font{\footnotesize \sffamily}
\makeatother

\title{The Price of Quantum Advantage in Computational Chemistry}

\author{Haejung Koh}
\email{h.koh@wisc.edu}
\affiliation{Department of Chemistry, University of Wisconsin--Madison, Madison, WI 53706, USA}

\author{Katerina Gratsea}
\affiliation{Department of Physics, University of Wisconsin--Madison, Madison, WI 53706, USA}

\author{Hao Zhang}
\affiliation{Department of Physics, University of Wisconsin--Madison, Madison, WI 53706, USA}

\author{Matthew Otten}
\email{mjotten@wisc.edu}
\affiliation{Department of Chemistry, University of Wisconsin--Madison, Madison, WI 53706, USA}
\affiliation{Department of Physics, University of Wisconsin--Madison, Madison, WI 53706, USA}

\begin{abstract}
As the fault-tolerant era approaches, the question shifts from ``is it possible'' to ``is it worthwhile.'' To assess practical quantum advantage against reliable classical baselines, we examine 26 benchmark instances including FeMoco, iron-sulfur clusters, cytochrome P450, and ruthenium catalysts using two selected configuration interaction solvers, SHCI and TrimCI.  Comparing classical workloads with fault-tolerant resource estimates, we calculate the break-even whole-QPU hourly price. Even at an optimistic price of US\$7{,}000 per whole-QPU hour, and with mid-range CPU clusters, classical computation remains cheaper for six of seven systems. FeMoco alone remains sensitive to the classical method and extrapolation strategy. Wavefunction analysis further reveals how orbital choice affects the determinant support required for target-state overlap in quantum phase estimation (QPE). Our results highlight the importance of reliable classical references and trial-state characterization, and quantify the economic conditions under which quantum computing can become competitive with classical alternatives.
\end{abstract}

\maketitle
\section{Introduction}
\label{sec: Intro}
Chemistry is considered one of the most promising applications for quantum computing~\cite{mcardle2020quantum,10.1021/acs.jpclett.6c00888}. Current quantum resource studies for chemistry have largely focused on ground-state energy estimation~\cite{PhysRevResearch.3.033055,nguyen2026corrosion,otten2024amyloid,watts2024propellant}. Faster energy estimation on fault-tolerant quantum computers may be attractive, but realizing such hardware requires substantial error correction and infrastructure overhead~\cite{Beverland2022QuantumResources,Mohseni:2024cay,otten2023qrechem}. Thus, classical methods remain preferable when they achieve the required accuracy at lower cost within the available time~\cite{lee2023evaluating,10.1021/acs.jpclett.6c00888}.

On the classical side, such comparisons require reliable classical baselines~\cite{chan2024spiers,bellonzi2025qb}. However, obtaining exact reference energies for large molecules remains challenging~\cite{Goings_2022,zhai2026classicalcomputationalsimulationfemocofactor,zhang2025randomdeterminantsgroundstate}. Strongly correlated systems such as FeMoco and iron-sulfur clusters, multiconfigurational cytochrome P450 (P450), and ruthenium-based catalysts dominated by dynamic correlation are in this regime~\cite{Goings_2022,sharma2014low,shaik2010p450,PhysRevResearch.3.033055,zhai2026classicalcomputationalsimulationfemocofactor}, while they play important roles in biological nitrogen fixation~\cite{doi:10.1073/pnas.1619152114,Li_2019}, electron transfer~\cite{beinert1997iron,sharma2014low}, oxidative drug metabolism~\cite{Guengerich2024,shaik2010p450}, and CO$_2$ hydrogenation~\cite{PhysRevResearch.3.033055}, respectively. Their active space Hamiltonians combine large determinant spaces, multiconfigurational wavefunctions, and dense low-energy spectra~\cite{Li_2019,sharma2014low,Goings_2022}, hindering accurate ground-state energy estimation. In P450 Compound~I (Cpd I), for example, near-degenerate doublet and quartet states underlie two-state reactivity~\cite{shaik2010p450,hirao2005two}, complicating energy and spin splitting validation~\cite{Goings_2022,sharma2014low}.

Among classical electronic-structure solvers for these problems, selected configuration interaction (SCI)~\cite{CIPSI_paper} occupies a distinctive position at the classical-quantum interface. SCI represents the target state, such as the ground state, as an explicit expansion over selected Slater determinants, each of which maps directly to a computational basis state under an occupation-number encoding~\cite{seeley2012bravyi}. The same expansion can therefore serve as a compact multireference trial state, a benchmark for validating quantum calculations, and a structured input for quantum state preparation. Its compressibility affects classical representation costs and, for a given scheme, the cost of preparing a high-overlap initial state~\cite{tubman2018postponingorthogonalitycatastropheefficient,feniou2024sparsequantumstatepreparation,Fomichev_2024,dcunha2024state}.

Other classical solvers probe different structures: density matrix renormalization group (DMRG)~\cite{PhysRevB.48.10345,PhysRevLett.69.2863,10.1063/1.1449459} approximates the wavefunction as a matrix-product state (MPS), whose bond dimension controls its capacity to represent entanglement~\cite{wouters2014density}. Single-reference coupled-cluster theory with single and double excitations and perturbative triple excitations (CCSD(T)) relies on its reference and truncated excitation hierarchy~\cite{bartlett2007coupled,bulik2015can}. These limitations make it difficult to establish the reliability of reference energies from a single method, particularly for strongly correlated systems. Agreement among independent methods with different approximations therefore provides greater confidence in the reference energies. 

On the quantum side, fault-tolerant quantum resource estimation (QRE) studies have quantified the resources required for catalytically relevant strongly correlated systems~\cite{doi:10.1073/pnas.1619152114,PhysRevResearch.3.033055,Goings_2022}. These studies translate logical-qubit and non-Clifford gate requirements into physical-qubit counts and runtimes under explicit hardware and error-correction assumptions. Comparisons of these estimates with classical costs have identified candidate regimes of quantum advantage~\cite{Goings_2022}, and utility-driven analyses have reported quantum and classical workloads in QPU-hours and CPU-hours~\cite{Goings_2022,gratsea2025achievingutilityscaleapplicationsstack,Beverland2022QuantumResources}. Because these units measure different resources, their ratio alone cannot determine which approach is faster or cheaper; we should consider classical concurrency and the price of each resource. As the classical algorithms improve, moreover, the boundaries of these regimes must be updated. Such assessments thus require reliable classical references not only for post hoc validation, but also for identifying genuinely difficult targets and verifying the intended low-energy state.

In this work, we use two SCI solvers, semistochastic heat-bath configuration interaction (SHCI)~\cite{Sharma_2017,Li_2018} and its recent variant, trimmed configuration interaction (TrimCI)~\cite{zhang2025randomdeterminantsgroundstate}, to establish classical baselines and assess the reliability of DMRG and SHCI references (Sec.~\ref{sec:IIIA}). For seven benchmarks, we compare classical runtimes and costs with fault-tolerant quantum resource estimates, quantifying two thresholds for quantum advantage: the classical concurrency required for time parity $\alpha_{\mathrm{eq}}$ and the break-even whole-QPU hourly price $\pi_{\mathrm{QPU}}^\star$ (Sec.~\ref{sec:IIIB}). We then examine how orbital choice and CI truncation affect trial-state overlaps and the determinant count needed to achieve a target overlap (Sec.~\ref{sec:IIIC}).

\section{Methods}
\label{sec: Method}

\subsection{Classical Methods: SHCI and TrimCI}
\label{sec:shci_trimci_formalism}

SHCI~\cite{Sharma_2017} approximates low-energy eigenstates by partitioning the full determinant space into a variational subspace $\mathcal{V}$ and a complementary perturbative space $\mathcal{P}$. The variational wavefunction is obtained by diagonalizing the Hamiltonian projected onto $\mathcal{V}$,
\begin{equation}
\ket{\Psi_{\mathcal{V}}}
=
\sum_{D_i \in \mathcal{V}}
c_i \ket{D_i}.
\label{eq:shci_wavefunction}
\end{equation}
The variational space is then expanded by adding determinants
$D_a \in \mathcal{P}$ satisfying the heat-bath selection criterion
\begin{equation}
\left| H_{ai} c_i \right| \ge \epsilon_1
\label{eq:shci_selection}
\end{equation}
for at least one determinant $D_i \in \mathcal{V}$~\cite{Sharma_2017}.
After each expansion, the Hamiltonian is rediagonalized within
$\mathcal{V}$, using a Davidson solver
\cite{DAVIDSON197587,DAVIDSON198949,MORGAN1992287}.

After convergence of the variational stage, the remaining correlation energy
is estimated using second-order Epstein--Nesbet perturbation theory~\cite{PhysRev.28.695,10.1098/rspa.1955.0134},
\begin{equation}
\Delta E_2 =
\sum_{D_a \in \mathcal{P}}
\frac{\left| \sum_{D_i \in \mathcal{V}} H_{ai} c_i \right|^2}
{E_0 - E_a},
\label{eq:EN_PT2}
\end{equation}
where $E_0$ is the variational energy and
$E_a=\langle D_a|H|D_a\rangle$. In practice, the perturbative correction is
screened using a threshold $\epsilon_2$ on the coupling terms
$|H_{ai}c_i|$. For small $\epsilon_2$, the number of perturbative
contributions becomes too large to evaluate fully deterministically, so SHCI
uses a semistochastic estimator that combines deterministic and stochastic
contributions~\cite{Sharma_2017}:
\begin{equation}
\Delta E_2(\epsilon_2) =
\big[
\Delta E^{\mathrm{s}}_2(\epsilon_2)
-
\Delta E^{\mathrm{s}}_2(\epsilon^{\mathrm{dtm}}_2)
\big]
+
\Delta E^{\mathrm{dtm}}_2(\epsilon^{\mathrm{dtm}}_2).
\label{eq:shci_pt2_total}
\end{equation}

In this work we use the fast semistochastic SHCI implementation
(FSHCI)~\cite{Li_2018}, which accelerates
Hamiltonian construction and evaluates the perturbative correction using
deterministic, pseudo-stochastic, and stochastic thresholds,
\begin{equation}
\epsilon^{\mathrm{dtm}}_2 >
\epsilon^{\mathrm{psto}}_2 >
\epsilon^{\mathrm{sto}}_2.
\end{equation}
Unless otherwise specified, we refer to this implementation as SHCI.

To complement SHCI, we use trimmed configuration interaction (TrimCI)~\cite{zhang2025randomdeterminantsgroundstate}. Whereas the SHCI expansion criterion expands $\mathcal{V}$ based only on Hamiltonian couplings to the current $\mathcal{V}$, TrimCI builds the determinant space through an
iterative expansion-trimming procedure. At each iteration, the determinant pool is expanded through Eq.~\eqref{eq:shci_selection} and then reduced by retaining determinants with the largest weights in the ground state of the pool. This additional trimming step consists of local truncation based on random grouping and global truncation, which ensure efficient memory control. It keeps the important and compact variational core, which corresponds to a heavily weighted and connected region in the Hamiltonian coupling graph. To reach energy uncertainties beyond the variational energy on the core, SHCI expansion and perturbation steps were further applied to the compact TrimCI variational core.

This distinction is useful for the present benchmarks because TrimCI provides a methodologically different route to the same true ground states. In particular, its amplitude-based trimming step can give a more balanced treatment of competing spin sectors when the dominant determinants differ between states. We therefore use TrimCI both as a compact SCI solver and as an independent check on SHCI and DMRG reference energies.

TrimCI’s iterative expansion and trimming can achieve comparable variational energies with fewer retained determinants than SHCI, making the final determinant count an incomplete measure of computational work. We therefore use SHCI trajectories to relate determinant count, residual PT2, and cost, referring to SHCI runs initialized from TrimCI spaces as TrimCI-seeded SHCI.

We prioritize variational energies in reference comparisons because they provide rigorous upper bounds to the exact ground-state energy of the specified Hamiltonian and symmetry sector. In several cases, variational energies alone approach available FCI values (Appendix Table~\ref{tab:trimci_best}) or fall below published ground-state estimates (Sec.~\ref{sec:IIIA}), establishing improved bounds without relying on perturbative corrections or extrapolation. We use PT2-corrected energies and extrapolations as complementary estimates, and the residual PT2 magnitude as a convergence diagnostic (Sec.~\ref{sec:IIIB}).

\subsection{Quantum Resource Estimation (QRE)}
\label{sec:QRE}

We performed quantum resource estimates for seven benchmark systems: Fe$_2$S$_2$, Fe$_4$S$_4$, FeMoco, P450 Cpd~I $2M_S=5$-x, and three Ru-based catalyst active spaces, selected as computationally demanding cases. Following the QRE methodology described in~\cite{gratsea2025achievingutilityscaleapplicationsstack}, ground-state energy estimation uses qubitization-based quantum phase estimation (QPE), employing double-factorized tensor hypercontraction (DFTHC), block-invariant symmetry shift (BLISS), and spectrum amplification (SA)~\cite{DFTHC}. The QRE workflow uses three open-source tools: \textsc{pyLIQTR}~\cite{Obenland_pyLIQTR}, \textsc{pyZX}~\cite{kissinger2020Pyzx}, and \textsc{Bench-Q}~\cite{benchq}. Proxy circuits were constructed to match the logical-qubit and $T$-gate counts in the work of~\cite{DFTHC}. An exception is Fe$_2$S$_2$, for which the double-factorized quantum algorithm implemented in \textsc{pyLIQTR}~\cite{DF_pyLIQTR} yields a lower $T$-gate count of $1\times10^{7}$. The workflow incorporates ZX-calculus-based circuit optimization and graph-state compilation to estimate the computational resources required for fault-tolerant execution.

Our resource estimates assume a distributed ion-trap architecture with surface-code error correction (cycle time of $10^{-4}\mathrm{s}$~\cite{Simon}) and an effective logical all-to-all connectivity. Magic-state distillation provides high-fidelity resource states for fault-tolerant $T$ gates and has been used as part of the quantum error correction scheme. QPE estimates include the block-encoding repetitions needed to achieve a target energy-estimation precision of $\sigma_{\mathrm{PEA}}=1$~mHa~\cite{gratsea2025achievingutilityscaleapplicationsstack}. In ideal QPE, the ground-state success probability is $\gamma=|\langle\Psi_{\mathrm{trial}}|\Psi_0\rangle|^2$, where $|\Psi_0\rangle$ denotes the exact ground state of the simulated Hamiltonian and $|\Psi_{\mathrm{trial}}\rangle$ is the approximate initial state supplied to QPE. Section~\ref{sec:IIIC} examines how orbital choice and CI truncation affect $\gamma$ and the retained determinant count $K$, a proxy for initial-state preparation complexity.

\section{Results}
\label{sec:results}

\subsection{Reference Validation}
\label{sec:IIIA}

\begin{figure}[htbp]
    \centering
    \includegraphics[width=1\linewidth]{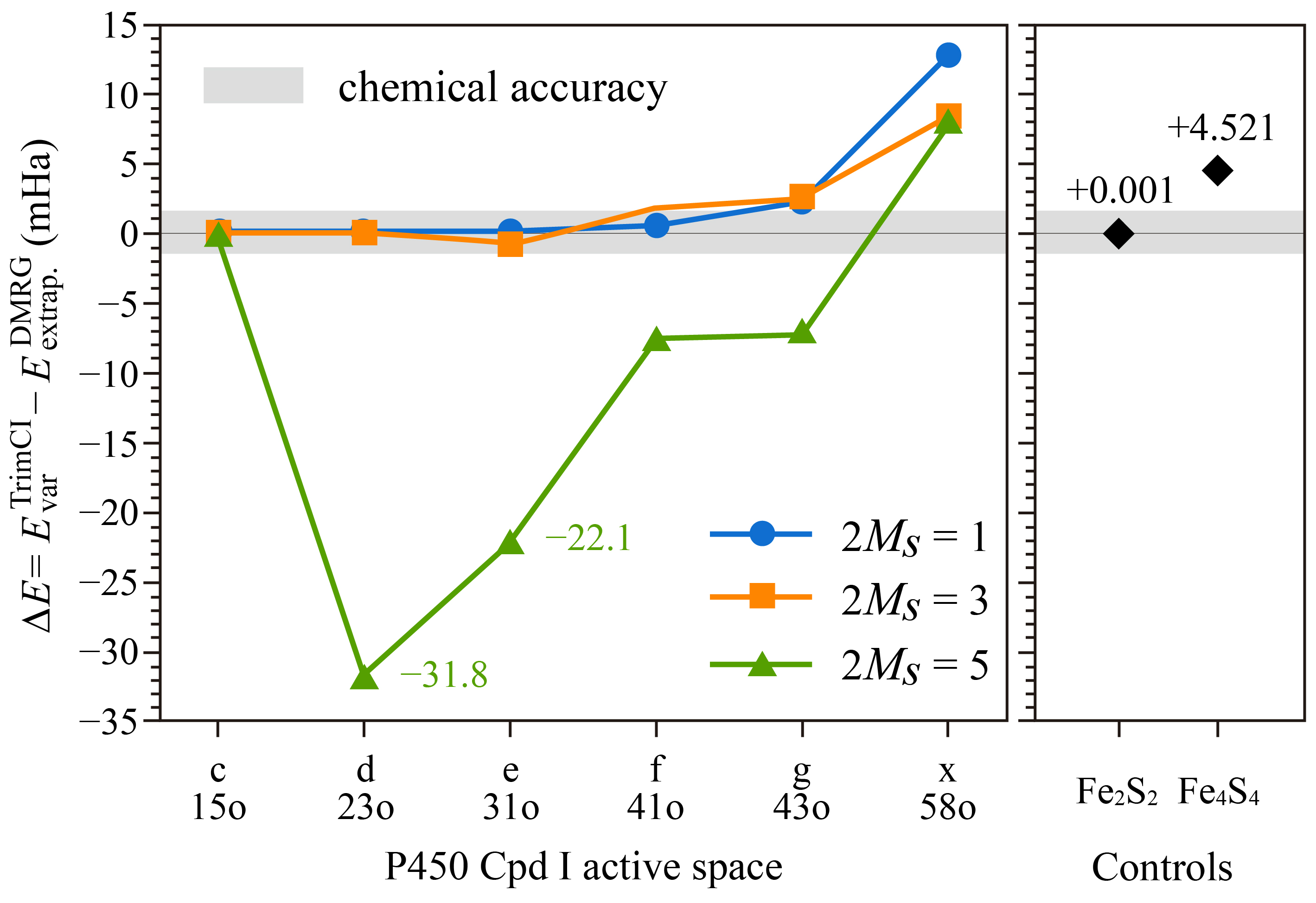}
    \caption{
    Comparison with published classical reference energies in~\cite{10.1021/acs.jctc.4c00903,Goings_2022}, using the variational energy of TrimCI-seeded SHCI. We report $\Delta E=E^{\mathrm{TrimCI}}_{\mathrm{var}}
    -E^{\mathrm{DMRG}}_{\mathrm{ref}}$ for P450 Compound~I active spaces, comparing the lowest energies obtained for the same Hamiltonian and $2M_S$. Because we report TrimCI-seeded SHCI variational energy, a negative $\Delta E$ establishes that the reference lies at least $|\Delta E|$ above the exact ground-state energy of the specified sector, irrespective of whether the reference itself is variational. The largest reference discrepancies occur in the $2M_S=5$ sector for active spaces $d$ ($31.8$~mHa) and $e$ ($22.1$~mHa), with $7.4$~mHa in $f$. Independent DMRG calculations reproduce the lower TrimCI bounds within $0.005$~mHa for $d$ and $0.27$~mHa for $e$, identifying the discrepancies as optimization-dependent reference errors. For Fe$_2$S$_2$ and Fe$_4$S$_4$, the TrimCI variational energies lie $0.001$ and $4.521$~mHa above the published references, respectively. The shaded gray region denotes chemical accuracy, $1.6$~mHa.
    }
\label{fig:solver-comparison}
\end{figure}

A meaningful classical cost requires a well-characterized classical reference. Over the past two decades, DMRG has become widely regarded as a benchmark standard for moderately and strongly correlated electronic systems~\cite{baiardi2020density}. We examine the published DMRG energies of P450~\cite{10.1021/acs.jctc.4c00903}, obtained for the same Hamiltonians in~\cite{Goings_2022}, against strictly variational TrimCI bounds and our own independent DMRG calculations detailed in Appendix~\ref{app:dmrg_support}.

Fig.~\ref{fig:solver-comparison} illustrates this comparison using the $2M_S=5$ sector as a representative example, showing the energy difference $\Delta E=E_{\mathrm{var}}^{\mathrm{TrimCI}}-E_{\mathrm{ref}}^{\mathrm{DMRG}}$. The shaded band indicates differences within $\delta=1.6$~mHa, approximately the conventional chemical-accuracy energy scale, which we adopt as a stopping diagnostic~\cite{10.1021/acs.jpca.8b01554}. Negative values indicate that TrimCI's variational energy alone, without PT2 correction, is lower than the published DMRG reference, thereby establishing a lower bound on the reference's error relative to the exact ground-state energy. In contrast, positive values indicate that TrimCI remains above the reference, without establishing how close the reference is to the exact energy.

The largest discrepancies are in P450 active spaces $d$ and $e$, where TrimCI variational energies lie $31.8$ and $22.1$~mHa below previously published DMRG references~\cite{10.1021/acs.jctc.4c00903} (Table~\ref{tab:trimci_best}). Our independent DMRG calculations agree with these lower energies within $0.5$~mHa. DMRG convergence to higher-energy local minima has long been recognized, particularly in connection with orbital ordering and initialization~\cite{RevModPhys.77.259,10.1063/1.4905329,10.1063/1.2139998}, and ground-state DMRG has also been observed to lock onto low-lying excited states~\cite{PhysRevResearch.5.023135}.

This gap is consistent with the ground-state and same-sector excitation energies recently reported in~\cite{Schroeder:2026zkz}. Our DMRG convergence analysis (Appendix Fig.~\ref{fig:dmrg-convergence}) further reveals that ground-state-targeted DMRG can converge to metastable higher-energy solutions, despite satisfying strict sweep-energy convergence criteria. Such misidentification can occur even in relatively modest active spaces, such as $d$ [25e, 23o] and $e$ [33e, 31o], where DMRG is expected to be accurate~\cite{wouters2014density}. Trial states extracted from the two DMRG branches differ in their squared overlaps with the SHCI reference by more than ten orders of magnitude at the same determinant count (Appendix~\ref{app:p450_state_validation}). Similar classical baseline concerns arise in SHCI, where reference-dependent determinant selection can lead to spurious spin-sector ordering under aggressive truncation (Appendix Fig.~\ref{fig:shci_trimCI_combined}).

\begin{figure*}[htbp]
    \centering
    \includegraphics[width=\linewidth]{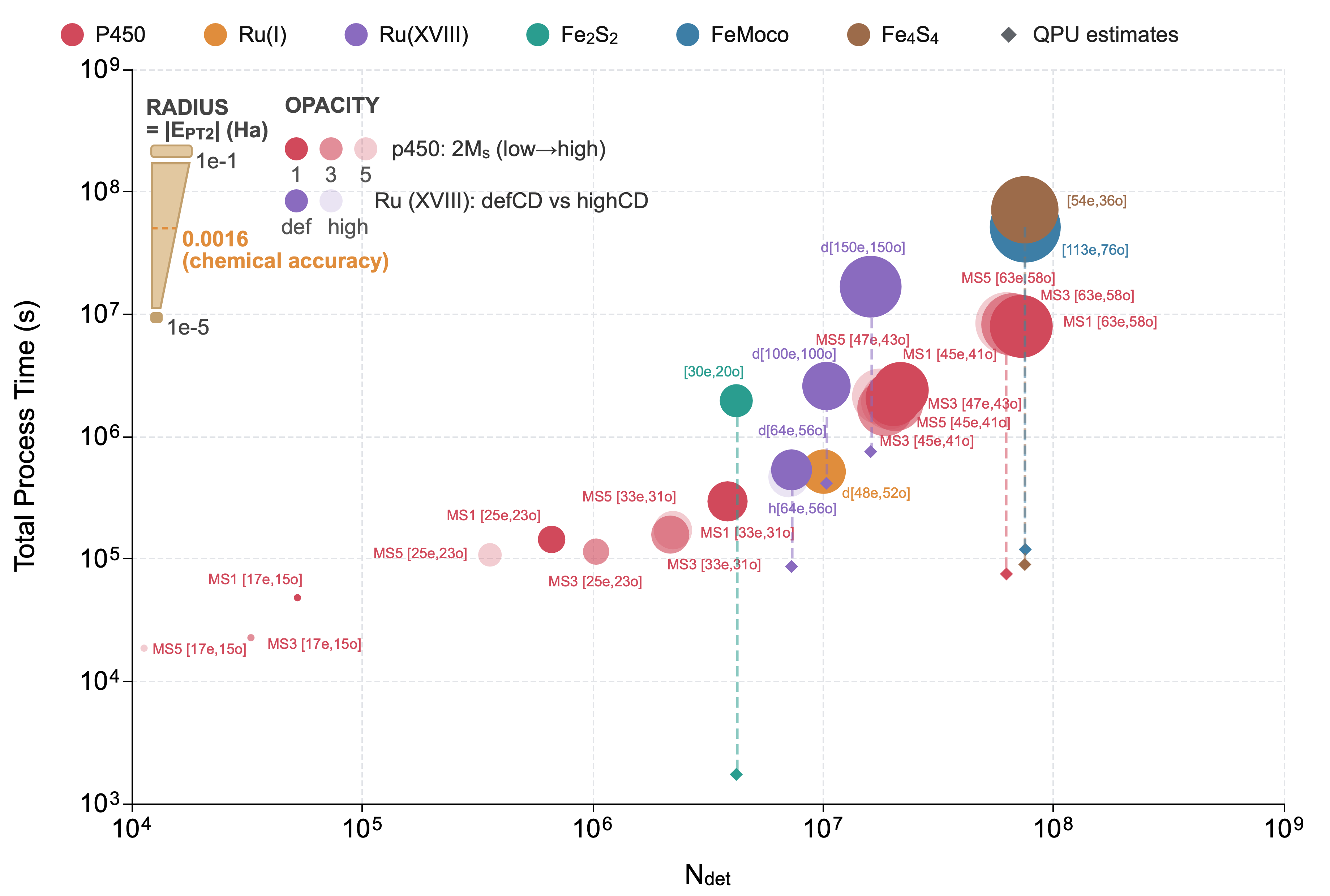}

    \vspace{0.1em}
    {\scriptsize
    \renewcommand{\arraystretch}{1.25}
    \begin{tabular*}{\textwidth}{@{\extracolsep{\fill}} ll rrr rr rrr @{}}
\toprule
& &
\multicolumn{5}{c}{Classical (SHCI)} &
\multicolumn{3}{c}{Quantum} \\
\cmidrule(lr){3-7}
\cmidrule(l){8-10}
System
& Active space
& $\epsilon_1$
& $|E_{\mathrm{PT2}}|$
& $N_{\mathrm{det}}$
& \multicolumn{2}{c}{$t_{\mathrm{CPU}}$}
& $N_{\mathrm{q}}^{\mathrm{basic}}$
& $N_{\mathrm{q}}^{\mathrm{detail}}$
& $t_{\mathrm{QPU}}$
\\
&
{\footnotesize [$N_e,N_o$]}
&
& (Ha)
&
& (core-s)
& (core-days)
& (M)
& (M)
& (days)
\\
\midrule
Fe$_2$S$_2$
& [30e, 20o]
& $5.00\times10^{-6}$
& $1.22\times10^{-4}$
& 4{,}208{,}596
& $1.96\times10^{6}$
& 22.66
& 1.30
& 4.28
& 0.02
\\
FeMoco
& [113e, 76o]
& $1.50\times10^{-5}$
& $5.13\times10^{-2}$
& 75{,}554{,}084
& $5.14\times10^{7}$
& $595.15^{*}$
& 6.10
& 15.70
& 1.38
\\
Fe$_4$S$_4$
& [54e, 36o]
& $7.00\times10^{-6}$
& $2.89\times10^{-2}$
& 75{,}238{,}321
& $7.12\times10^{7}$
& $823.53^{*}$
& 2.29
& 7.55
& 1.04
\\
P450-x
& [63e, 58o] ($2M_S=5$)
& $1.00\times10^{-5}$
& $1.26\times10^{-2}$
& 62{,}610{,}882
& $8.42\times10^{6}$
& $97.48^{*}$
& 4.66
& 14.80
& 0.87
\\
\midrule
\multirow{3}{*}{Ru(XVIII)}
& [150e, 150o]
& $1.00\times10^{-5}$
& $1.18\times10^{-2}$
& 16{,}124{,}840
& $1.67\times10^{7}$
& $193.84^{*}$
& 5.70
& 18.10
& 8.70
\\
&
[100e, 100o]
& $5.00\times10^{-6}$
& $1.40\times10^{-3}$
& 10{,}365{,}091
& $2.58\times10^{6}$
& $29.89$
& 3.70
& 9.40
& 4.80
\\
&
[64e, 56o]
& $5.00\times10^{-6}$
& $4.33\times10^{-4}$
& 7{,}313{,}547
& $5.33\times10^{5}$
& 6.17
& 1.80
& 5.40
& 1.00
\\
\bottomrule
\addlinespace[1pt]
\multicolumn{10}{r}{%
\scriptsize
$^{*}$Classically non-converged;
$t_{\mathrm{CPU}}$ is a lower bound.%
}
\\
\end{tabular*}
    }
    \caption{Measured SHCI resources and paired fault-tolerant quantum-resource estimates. The scatter plot shows total CPU time, including variational optimization and PT2 evaluation, versus $N_{\mathrm{det}}$ for each reported calculation. Bubble size denotes $|E_{\mathrm{PT2}}|$. Diamonds give the expected QPU runtime $t_{\mathrm{QPU}}$ of each paired system, placed at the $N_{\mathrm{det}}$ of the corresponding SHCI final runs for visual pairing; their horizontal position carries no quantum-resource information. For each of the seven representative systems, the table reports the smallest selection threshold $\epsilon_1$, the corresponding PT2 correction and determinant count, total CPU time (including seed generation), estimated physical-qubit requirements, and projected QPU runtime. $N_{\mathrm{q}}^{\mathrm{basic}}$ denotes the physical-qubit count obtained from the basic surface-code footprint model, which includes logical data qubits and magic-state factories but does not take into account the number of ions that each Elementary Logical Unit (ELU) needs to support the $10^{-4}\mathrm{s}$ surface cycle time at the inter-ELU level~\cite{Simon}. In contrast, $N_{\mathrm{q}}^{\mathrm{detail}}$ is obtained from a detailed modular ion-trap hardware model that accounts for the communication, computation, and memory ions required within each ELU to support inter-ELU operations at the assumed surface-code cycle time~\cite{gratsea2025achievingutilityscaleapplicationsstack, Simon}.
    We take classical core times from the Portable Batch System (PBS) accounting. All quantum-resource estimates target an energy precision of $1$~mHa and use the hardware model and surface-code cycle time of Sec.~\ref{sec:QRE}. For FeMoco, Fe$_4$S$_4$, P450-x, and Ru(XVIII)[150e, 150o], SHCI final runs remain unconverged to the chemical-accuracy threshold of $1.6$~mHa. Thus, their recorded work is a lower bound on the cumulative work needed to reach the target by continuing the same workflow. Ru(I) and Ru(XVIII) refer to structures I and XVIII, respectively, in the ruthenium-catalyzed CO$_2$ hydrogenation cycle studied in~\cite{PhysRevResearch.3.033055}. Additional data are given in Table~\ref{tab:allinone}.}
    \label{fig:combined}
\end{figure*}

\subsection{Classical-Quantum Resource Comparison}
\label{sec:IIIB}

Fig.~\ref{fig:combined} presents the total CPU times (bubbles) and QPU times (diamonds) as a function of $N_{\mathrm{det}}$ at the final variational step. Bubble radii encode the magnitude of the residual PT2 correction, $|E_{\mathrm{PT2}}|$, providing an internal convergence diagnostic. Among our systems, FeMoco and Fe$_4$S$_4$ exhibit the two largest bubble radii, indicating substantial $|E_{\mathrm{PT2}}|$ in our best trials and suggesting that they are our most difficult benchmarks. P450 generally exhibits intermediate $|E_{\mathrm{PT2}}|$ among our benchmarks, with larger corrections tending to occur in larger active spaces and in the $2M_S=1$ and $3$ sectors compared with $2M_S=5$. Ru instances with $\leq 100$ electrons also have PT2 corrections within chemical accuracy. These differing levels of convergence motivate comparing classical resource requirements at a common residual-PT2 threshold. For seven benchmark systems, QRE assumes $1$~mHa precision, indicated by uniformly sized diamonds. In all seven, the numerical QPU runtimes are smaller than the CPU times. 

\begin{figure*}[htbp]
\centering
\includegraphics[width=\linewidth]{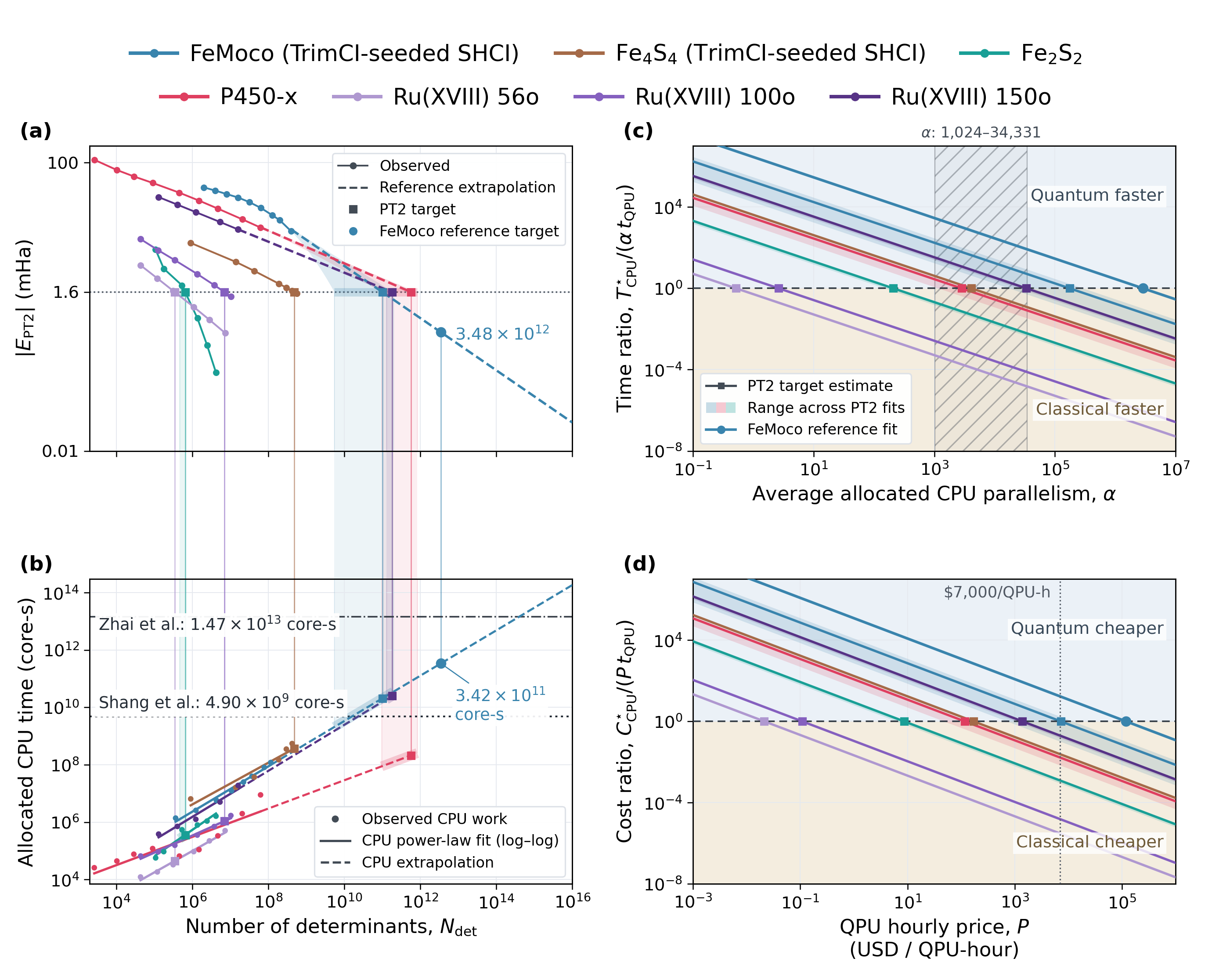}
{\scriptsize
\renewcommand{\arraystretch}{1.25}
\setlength{\tabcolsep}{4pt}
\begin{tabular*}{\linewidth}{@{\extracolsep{\fill}} ll rrrrrr @{}}
\toprule
& &
\multicolumn{2}{c}{(a)} &
\multicolumn{1}{c}{(b)} &
\multicolumn{2}{c}{(c)} &
\multicolumn{1}{c}{(d)} \\
\cmidrule(lr){3-4}
\cmidrule(lr){5-5}
\cmidrule(lr){6-7}
\cmidrule(lr){8-8}
System
& Criterion
& $N_{\mathrm{det}}^{\star}$
& $|E_{\mathrm{PT2}}|$
& $t_{\mathrm{CPU}}^{\mathrm{alloc},\star}$
& $t_{\mathrm{QPU}}$
& $\alpha_{\mathrm{eq}}$
& $\pi_{\mathrm{QPU}}^{\star}$
\\
&
&
&
(mHa)
& (core-s)
& (days)
&
& (US\$ per whole-QPU hour)
\\
\midrule
FeMoco
& PT2
& $(5.477\text{--}194.5)\times10^{9}$
& $1.6$
& $(2.084\text{--}35.02)\times10^{9}$
& $1.38$
& $(1.748\text{--}29.37)\times10^{4}$
& $721$--$12{,}114$
\\
Fe$_4$S$_4$
& PT2
& $(4.814\text{--}4.923)\times10^{8}$
& $1.6$
& $(3.691\text{--}3.752)\times10^{8}$
& $1.04$
& $(4.108\text{--}4.176)\times10^{3}$
& $169.46$--$172.25$
\\
Fe$_2$S$_2$
& PT2
& $(4.593\text{--}6.690)\times10^{5}$
& $1.6$
& $(2.624\text{--}3.706)\times10^{5}$
& $0.02$
& $151.9$--$214.5$
& $6.265$--$8.846$
\\
P450-x
& PT2
& $(9.583\text{--}83.34)\times10^{10}$
& $1.6$
& $(0.8808\text{--}2.561)\times10^{8}$
& $0.87$
& $(1.172\text{--}3.407)\times10^{3}$
& $48.34$--$140.53$
\\
Ru(XVIII) 56o
& PT2
& $(3.412\text{--}3.514)\times10^{5}$
& $1.6$
& $(4.414\text{--}4.512)\times10^{4}$
& $1.00$
& $0.5108$--$0.5222$
& $0.02107$--$0.02154$
\\
Ru(XVIII) 100o
& PT2
& $(7.063\text{--}7.263)\times10^{6}$
& $1.6$
& $(1.090\text{--}1.108)\times10^{6}$
& $4.8$
& $2.629$--$2.673$
& $0.10843$--$0.11026$
\\
Ru(XVIII) 150o
& PT2
& $(1.268\text{--}1.812)\times10^{11}$
& $1.6$
& $(1.900\text{--}2.529)\times10^{10}$
& $8.7$
& $(2.527\text{--}3.365)\times10^{4}$
& $1{,}043$--$1{,}388$
\\
\midrule
FeMoco
& $E_{\mathrm{var}}$, log fit
& $3.481\times10^{12}$
& $0.4436$
& $3.423\times10^{11}$
& $1.38$
& $2.871\times10^{6}$
& $1.184\times10^{5}$
\\
\bottomrule
\end{tabular*}
}
\caption{
Classical resource projections and conditions for quantum time and cost advantages.
PT2 ranges span the minimum and maximum estimates from the power-law and log-quadratic fits (Appendix~\ref{app:pt2-sensitivity}). \textbf{(a)} Residual PT2 magnitude versus determinant count; filled squares indicate the $1.6$~mHa targets, interpolated for Fe$_4$S$_4$, Fe$_2$S$_2$, Ru(XVIII) 56o, and Ru(XVIII) 100o and extrapolated otherwise. Fe$_4$S$_4$ and FeMoco use TrimCI-seeded SHCI, which differs from Hartree--Fock seeded SHCI in Fig.~\ref{fig:combined}. The large FeMoco circle denotes the logarithmic variational-energy target, projected onto the PT2 curve (Appendix Fig.~\ref{fig:femoco-projection}). \textbf{(b)} Allocated CPU work ranges estimated by ordinary least-squares fits in log--log space to all available SHCI workload points, with the intercept allowed to vary. For FeMoco, the black dash-dotted line indicates the published two-sweep spin-unrestricted DMRG (UDMRG) cost estimate~\cite{zhai2026classicalcomputationalsimulationfemocofactor}, and the gray dotted line indicates the allocated CPU work inferred from the reported QiankunNet+AFQMC runtime~\cite{shang2026closing} (1,280 nodes, 608 cores per node, 1.75~h; $\sim 4.9\times10^{9}$ core-s; calculated here). \textbf{(c)} Classical-to-quantum execution-time ratio versus average allocated CPU cores, $\alpha$, with equality thresholds $\alpha_{\mathrm{eq}}=t_{\mathrm{CPU}}^{\mathrm{alloc},\star}/t_{\mathrm{QPU}}$, using consistent time units. The vertical band shows observed average allocations; QPU runtimes are the same as in Fig.~\ref{fig:combined}. \textbf{(d)} Classical-to-quantum compute-cost ratio versus  whole-QPU hourly price, $\pi_{\mathrm{QPU}}$ (axis label $P$),  at US\$0.04125/CPU core-hour. Break-even prices satisfy  $\pi_{\mathrm{QPU}}^{\star}=\pi_{\mathrm{CPU}}\alpha_{\mathrm{eq}}$. The US\$7{,}000/h reference exceeds all non-FeMoco ranges  but falls within the FeMoco PT2 range. Blue curves and circles denote the FeMoco variational-energy  log-fit scenario. Squares in (c,d) mark PT2-based time and  cost parity; ratios above unity favor quantum computation.
}
\label{fig:quantum-gain}
\end{figure*}

However, quantum estimates here assume a single QPU, whereas classical computations run across multiple nodes. Classical parallelism and relative hardware costs can substantially affect assessments of quantum advantage~\cite{Gundlach:2025sir}. The time advantages should therefore be interpreted in the context of these resource assumptions. In addition, comparing costs requires the hourly price of each computing resource: a faster quantum calculation may still be more expensive if QPU access costs more. We therefore distinguish allocated CPU work from execution concurrency and introduce the whole-QPU hourly price, defined as the charge for one hour of exclusive access to the entire fault-tolerant QPU required for the calculation. Moreover, our SHCI calculations terminate at different levels of convergence, so their CPU workloads cannot be directly compared at a common convergence threshold. We therefore estimate the determinant counts required to reach this threshold, assess their sensitivity to the assumed PT2 convergence model, and translate them into allocated CPU work using molecule-specific scaling. Fig.~\ref{fig:quantum-gain} examines how this price and CPU concurrency determine when quantum computation becomes faster or cheaper, using these expected classical workloads.

In Fig.~\ref{fig:quantum-gain}(a), we estimate the determinant count required to reach a residual-PT2 threshold of $\delta=1.6$~mHa. For Fe$_2$S$_2$, Fe$_4$S$_4$, Ru(XVIII)[64e, 56o], and Ru(XVIII)[100e, 100o], they already attain $\delta$, we only use log--log interpolation between the adjacent observations surrounding the threshold. For those remaining systems, P450-x, FeMoco,  and Ru(XVIII)[150e, 150o] we fit a power law to the last five PT2 observations:
\begin{equation}
\begin{aligned}
|E_{\mathrm{PT2}}| &= A N_{\mathrm{det}}^{-a},\\
N_{\mathrm{det}}^{\star} &= \left(\frac{A}{\delta}\right)^{1/a},
\end{aligned}
\label{eq:pt2_target_projection}
\end{equation}
where $A$ is an energy-valued prefactor and $a>0$ is the fitted decay exponent. We further assess convergence-model sensitivity using power-law and log-quadratic fits to the last 4, 5, or 6 observations, where available. Model definitions, fitting procedures, and root-selection criteria are detailed in Appendix~\ref{app:pt2-sensitivity}.

For FeMoco, we also estimate the determinant count required to reach the variational-energy target of $-22140.408505$~Ha, taken from a study of classical FeMoco simulation to chemical accuracy~\cite{zhai2026classicalcomputationalsimulationfemocofactor}. This provides a variational-energy criterion to the residual-PT2 threshold for FeMoco. The resulting projections are included in Fig.~\ref{fig:quantum-gain}(a,b), with fitting details in Appendix~\ref{sec:femoco-estimation}.

Fig.~\ref{fig:quantum-gain}(b) converts these determinant target counts into allocated classical work from our SHCI data. We denote this work by $t_{\mathrm{CPU}}^{\mathrm{alloc}}$, measured in core-sec, to distinguish it from the CPU time $t_{\mathrm{CPU}}$ obtained from PBS accounting in Fig.~\ref{fig:combined}. We use allocated work because actual CPU usage cannot be determined precisely at intermediate points along the trajectories. This work sums the reserved physical cores multiplied by elapsed time for each included step: initial setup, cumulative variational work, and the PT2 stage. This yields a conservative classical cost estimate, penalizing the classical baseline relative to utilization-based accounting.

The projected allocated work is
\begin{equation}
t_{\mathrm{CPU}}^{\mathrm{alloc},\star}
=
t_{\mathrm{CPU},0}^{\mathrm{alloc}}
\left(
\frac{N_{\mathrm{det}}^{\star}}{N_0}
\right)^b,
\label{eq:target_classical_work}
\end{equation}
where \(N_0=\exp[\langle\ln N_{\mathrm{det}}\rangle]\) is the
geometric mean of the determinant counts included in the timing fit.
For each system, \(t_{\mathrm{CPU},0}^{\mathrm{alloc}}\) and \(b\)
are obtained by ordinary least-squares fitting in log--log space to
all available workload points, with the intercept allowed to vary. Fe$_4$S$_4$ and FeMoco use TrimCI-seeded SHCI trajectories, while the other systems start from the Hartree--Fock determinant according to the original SHCI step. With the work model held fixed, the power-law PT2 model gives
$t_{\mathrm{CPU}}^{\mathrm{alloc},\star}\propto(A/\delta)^{b/a}$,
showing how slow convergence can amplify the workload required to tighten the threshold. The FeMoco study cited above projects $4.08\times10^{9}$ CPU core-hours ($1.47\times10^{13}$ core-s) for two UDMRG sweeps at $D\approx393{,}000$, extrapolated from measured UDMRG timings to obtain a variational UDMRG energy within chemical accuracy of its extrapolated ground-state estimate~\cite{zhai2026classicalcomputationalsimulationfemocofactor}.

Fig.~\ref{fig:quantum-gain}(c) compares execution times as a function of classical concurrency. We define
$\alpha=t_{\mathrm{CPU}}^{\mathrm{alloc}}/t_{\mathrm{CPU,wall}}$
as the average number of allocated cores, where
$t_{\mathrm{CPU,wall}}$ is the elapsed time over the same SHCI intervals included in the work accounting. The quantum runtime $t_{\mathrm{QPU}}$ is the execution time of one modeled fault-tolerant QPU for the same active-space Hamiltonian, under the assumptions of Sec.~\ref{sec:QRE}. For each workload, the classical-to-quantum runtime ratio is
\begin{equation}
\begin{aligned}
R_t(\alpha)
&=
\frac{t_{\mathrm{CPU}}^{\mathrm{alloc},\star}}
{\alpha t_{\mathrm{QPU}}}
=
\frac{\alpha_{\mathrm{eq}}}{\alpha},\\
\alpha_{\mathrm{eq}}
&=
\frac{t_{\mathrm{CPU}}^{\mathrm{alloc},\star}}
{t_{\mathrm{QPU}}}.
\end{aligned}
\label{eq:target_runtime_ratio}
\end{equation}
Here $\alpha_{\mathrm{eq}}$ is the concurrency required for equal execution times, and $R_t>1$ indicates a shorter modeled quantum runtime. This comparison holds allocated work fixed as concurrency varies. In practice, communication overhead and idle capacity can change the allocated work, so the inferred concurrency requirements are conditional on this work model.

Fig.~\ref{fig:quantum-gain}(d) compares compute costs as a function of the whole-QPU hourly price. Let $\pi_{\mathrm{CPU}}$ denote the price per allocated CPU core-hour and $\pi_{\mathrm{QPU}}$ the price per whole-QPU hour. With allocated CPU work in core-hours and QPU runtime in hours, the cost model is
\begin{equation}
C_{\mathrm{CPU}}=\pi_{\mathrm{CPU}}t_{\mathrm{CPU}}^{\mathrm{alloc}},
\qquad
C_{\mathrm{QPU}}=\pi_{\mathrm{QPU}}t_{\mathrm{QPU}},
\label{eq:compute_price_model}
\end{equation}
where $C_{\mathrm{CPU}}$ and $C_{\mathrm{QPU}}$ are the classical and quantum compute costs, respectively.

We use $\pi_{\mathrm{CPU}}=\mathrm{US\$}\,0.04125$ per core-hour, based on the Linux On-Demand rate of US\$7.92 per hour for a 192-core EC2 \texttt{hpc8a.96xlarge} instance in the US East (Ohio) region~\cite{AWSEC2OnDemandPricing2026,AWSHpc8a2026}. For quantum execution, we refer to a price of $\pi_{\mathrm{QPU}}=\mathrm{US\$}\,7{,}000$ per whole-QPU hour, based on dedicated reservations of IonQ's 36-qubit Forte through Amazon Braket Direct~\cite{AWSBraketPricing2026}. This optimistically assumes that the fault-tolerant QPUs with millions of physical qubits required by our resource estimates in Fig.~\ref{fig:combined} are available at the same whole-machine hourly price. Further pricing details are documented in Appendix Table~\ref{tab:illustrative_cloud_pricing}.

For each projected classical workload, the break-even QPU hourly price is
\begin{equation}
\pi_{\mathrm{QPU}}^{\star}
=
\frac{C_{\mathrm{CPU}}^{\star}}{t_{\mathrm{QPU}}}
=
\frac{\pi_{\mathrm{CPU}}t_{\mathrm{CPU}}^{\mathrm{alloc},\star}}
{t_{\mathrm{QPU}}}.
\label{eq:target_break_even}
\end{equation}
The classical-to-quantum cost ratio is therefore
\begin{equation}
R_C
=
\frac{C_{\mathrm{CPU}}^{\star}}{C_{\mathrm{QPU}}}
=
\frac{\pi_{\mathrm{QPU}}^\star}{\pi_{\mathrm{QPU}}}.
\end{equation}
Thus, $R_C=1$ indicates equal costs, and quantum computation has the lower modeled cost when $\pi_{\mathrm{QPU}}<\pi_{\mathrm{QPU}}^\star$. When $R_C<1$, its reciprocal gives the factor by which classical computation is cheaper.

At the reference QPU price, classical computation remains cheaper for Fe$_2$S$_2$, Fe$_4$S$_4$, P450-x, and all three Ru(XVIII) active spaces. Even the highest break-even price among these six systems, US\$1{,}388 per QPU-hour for Ru(XVIII)[150e, 150o], is approximately fivefold below the US\$7{,}000 reference rate. For Fe$_2$S$_2$, Ru(XVIII)-56o, and Ru(XVIII)-100o, the upper bounds of the break-even price range are more than two orders of magnitude below this rate. The relative tractability of the Ru systems with our classical method therefore limits quantum cost competitiveness to substantially lower whole-QPU hourly prices.

Notably, Fe$_4$S$_4$, with only 36 active orbitals, has a break-even whole-QPU price of approximately US\$170 per hour, whereas Ru(XVIII)-100o, with 100 active orbitals, has a threshold of approximately US\$0.11 per hour (Fig.~\ref{fig:quantum-gain}(d)). This contrast illustrates that size-based forecasts, which place the quantum--classical crossover by system size and asymptotic algorithmic scaling~\cite{Gundlach:2025sir}, are insufficient to predict quantum economic competitiveness, which depends on the instance-specific classical workload required to reach the target convergence criterion as well as the quantum execution time.

FeMoco's break-even prices range from US\$721 to 12{,}114 per whole-QPU hour. This range includes the reference price of US\$7{,}000 per whole-QPU hour, so which approach is cheaper depends on the classical extrapolation strategy. The separate variational-energy projections favor quantum computation under the same pricing assumption (Appendix Fig.~\ref{fig:femoco-projection}). Panels (c) and (d) distinguish runtime advantage from cost advantage: quantum computation is faster while classical computation remains cheaper when $\alpha<\alpha_{\mathrm{eq}}$ and $\pi_{\mathrm{QPU}}>\pi_{\mathrm{QPU}}^\star$.

This comparison is inherently asymmetric. A single sufficiently accurate classical method can establish classical cost competitiveness with quantum computing, whereas demonstrating a quantum advantage requires comparison with the best available classical solution. Thus, our SHCI-based results here establish competitive classical costs for six systems under the stated assumptions, while the potential quantum advantage for FeMoco remains sensitive to ongoing advances in classical computational methods and assumptions about future quantum computing costs.

\begin{figure*}[htbp]
    \centering

    \begin{minipage}[t]{0.49\linewidth}
        \centering
        \textbf{(a)} Canonical HF basis\\[0.4em]
        \includegraphics[width=\linewidth]{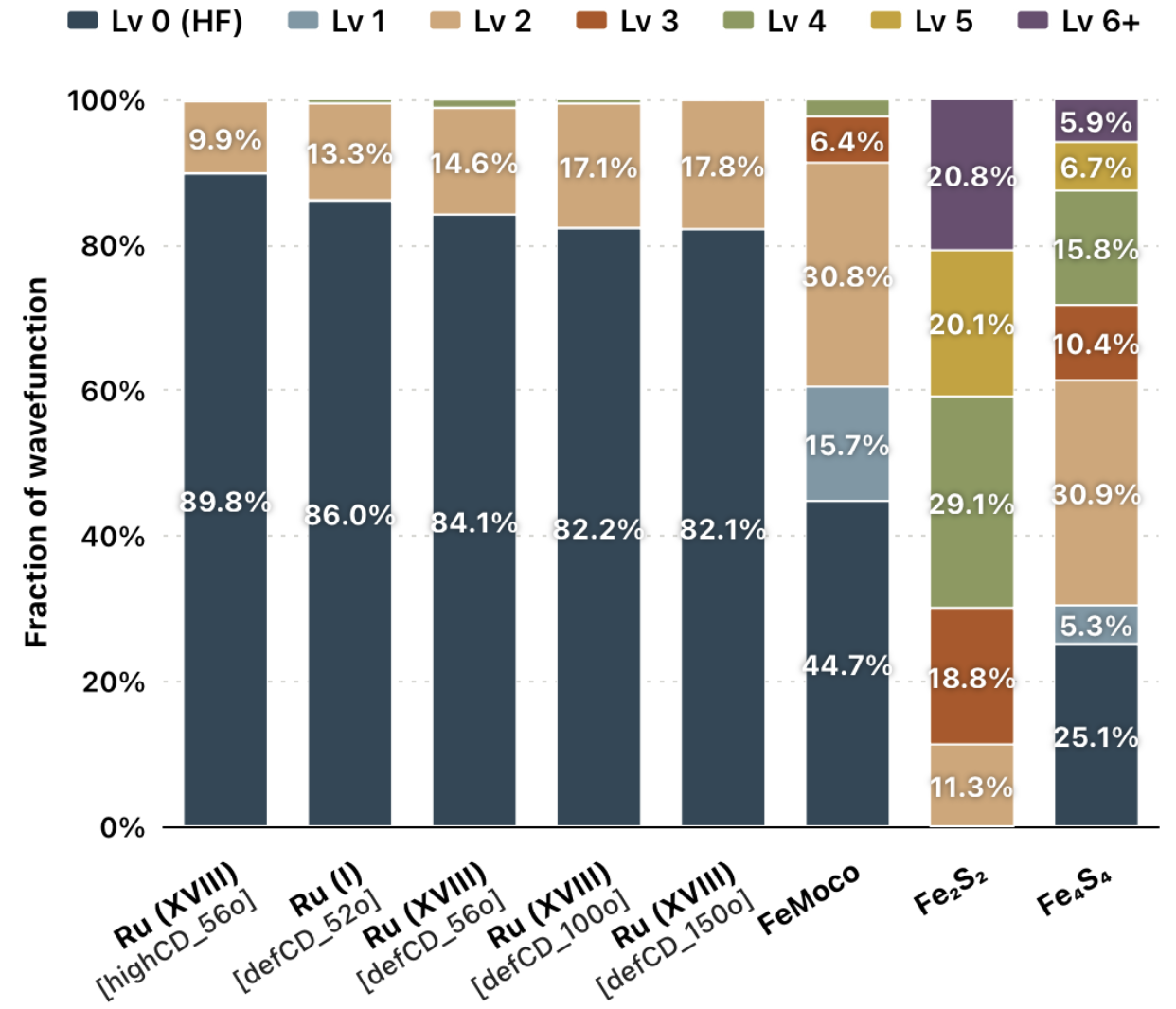}
    \end{minipage}
    \hfill
    \begin{minipage}[t]{0.49\linewidth}
        \centering
        \textbf{(b)} Energy-optimized basis\\[0.4em]
        \includegraphics[width=\linewidth]{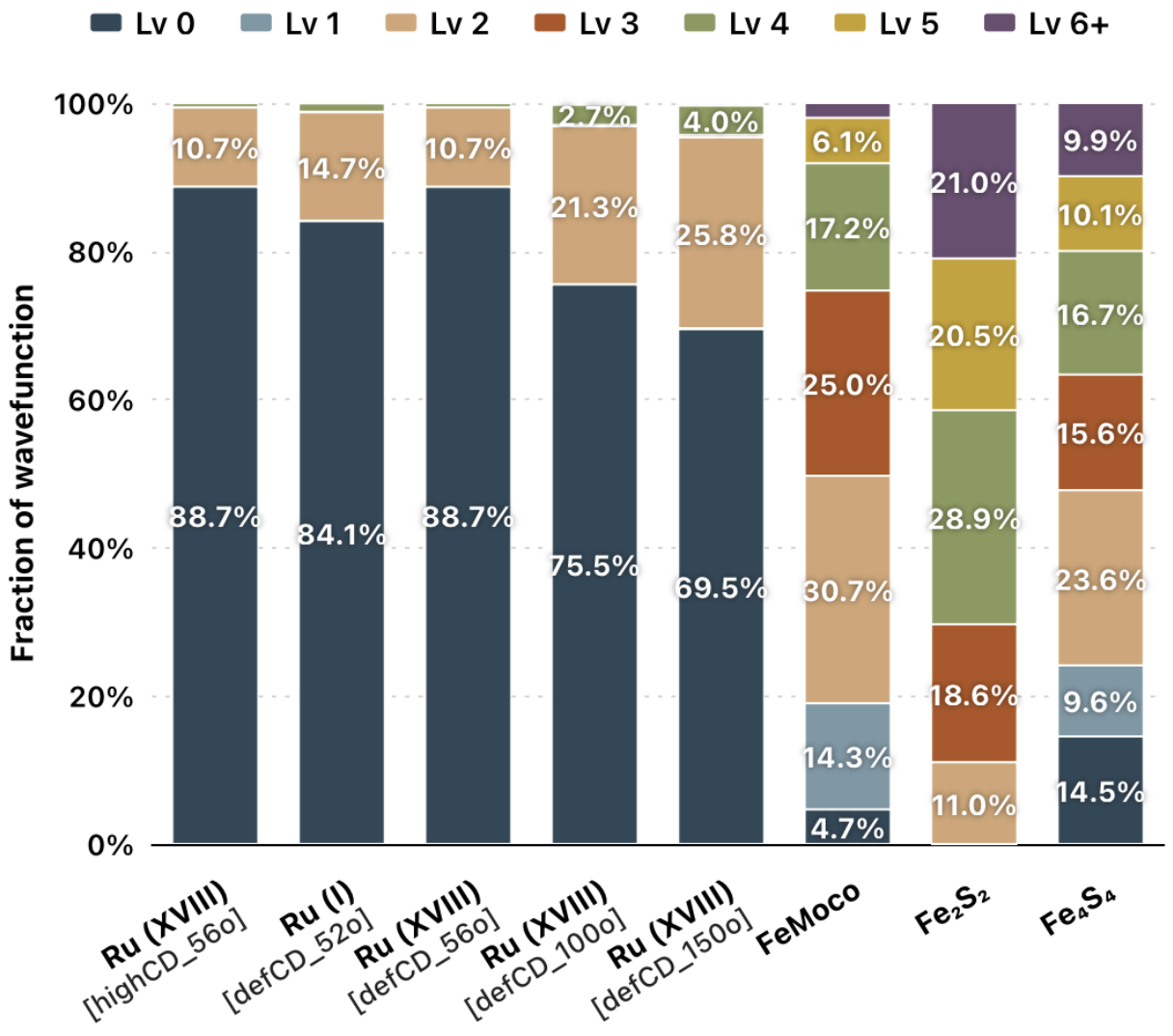}
    \end{minipage}
    \par\vspace{1.0em}

    \begingroup
    \small
    \setlength{\tabcolsep}{10pt}
    \renewcommand{\arraystretch}{1.1}
    \begin{tabular}{@{}lcc@{}}
        \multicolumn{3}{@{}l@{}}{\textbf{(c)}}\\[0.4em]
        \toprule
        System & Active space &
        $L_1=\sum_L|W_L^{\mathrm{HF}}-W_L^{\mathrm{OPT}}|$ \\
        \midrule
        FeMoco             & [113e, 76o]   & 0.8314 \\
        Fe$_4$S$_4$        & [54e, 36o]    & 0.3575 \\
        Ru(XVIII) defCD    & [150e, 150o] & 0.2523 \\
        Ru(XVIII) defCD    & [100e, 100o] & 0.1351 \\
        Ru(XVIII) defCD    & [64e, 56o]   & 0.0914 \\
        Ru(I) defCD        & [48e, 52o]   & 0.0408 \\
        Ru(XVIII) highCD   & [64e, 56o]   & 0.0219 \\
        Fe$_2$S$_2$        & [30e, 20o]   & 0.0113 \\
        \bottomrule
    \end{tabular}
    \par
    \endgroup
    \caption{
    Excitation-level decomposition of the SHCI variational
    states for eight transition-metal systems in
    \textbf{(a)} canonical Hartree--Fock basis and
    \textbf{(b)} an energy-optimized orbital basis.
    Each stacked bar gives the weight $W_L$ carried by determinants
    at excitation level $L$ relative to the reference occupation
    pattern in the corresponding basis.
    \textbf{(c)} Differences between the two profiles, quantified by
    $L_1=\sum_L|W_L^{\mathrm{HF}}-W_L^{\mathrm{OPT}}|$.
    For normalized distributions, $0\leq L_1\leq2$, with larger values indicating greater differences between excitation profiles, as observed in FeMoco. The labels defCD and highCD denote the Cholesky-decomposition thresholds used to construct the active-space Hamiltonians, with values of $10^{-4}$ (default) and $10^{-8}$
    (tighter), respectively. A small $L_1$ indicates similar profiles, but does not imply Hartree--Fock dominance or a compact determinant expansion,
    as illustrated by Fe$_2$S$_2$. The profiles identify excitation support relevant to initial-state selection; preparation costs also depend on the determinant amplitudes and the chosen circuit.
    }
    \label{fig:excited_lev}
\end{figure*}

\subsection{Wavefunction Analysis}
\label{sec:IIIC}

\begin{figure}[htbp]
    \centering
    \includegraphics[width=1\linewidth]{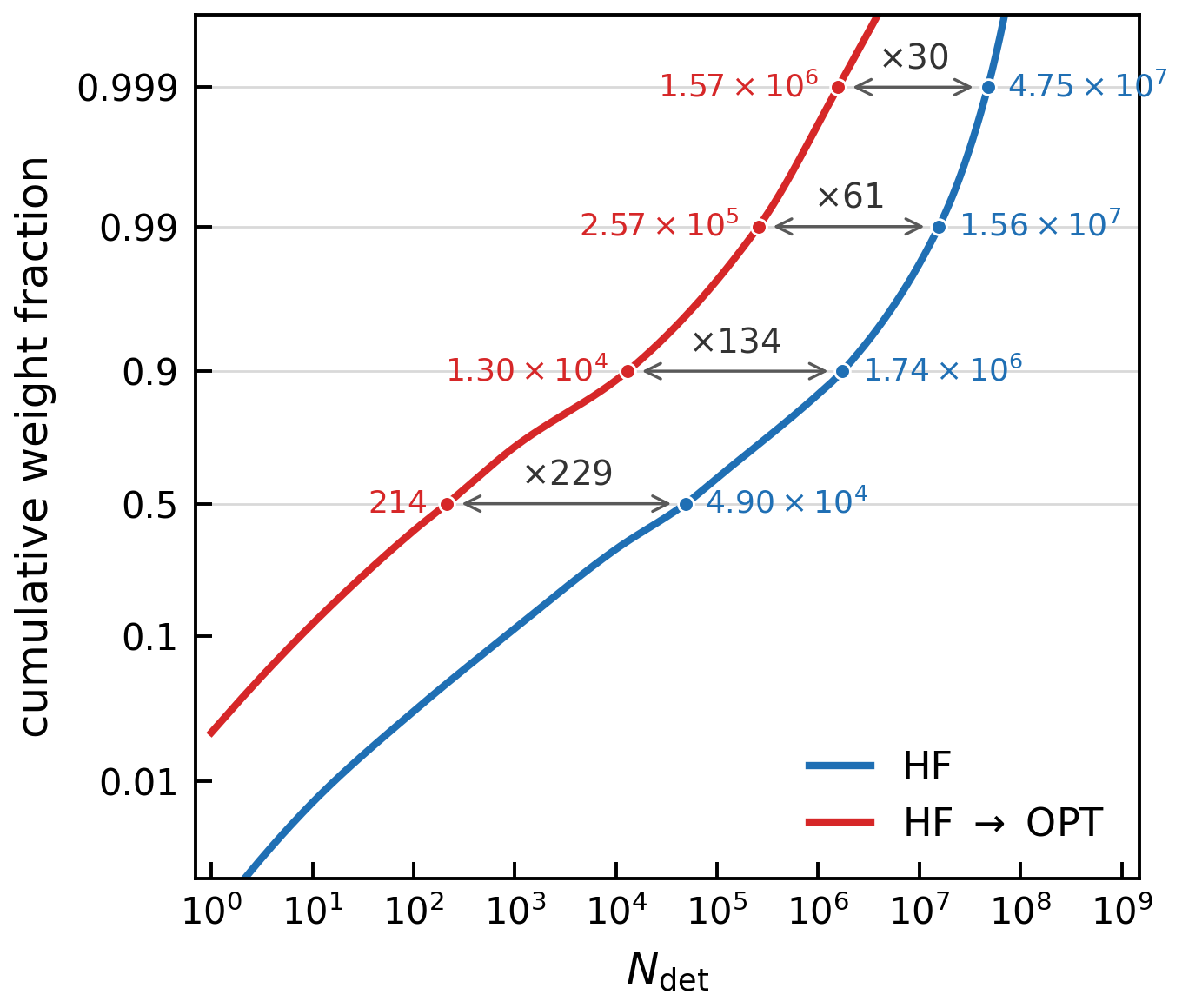}
    \caption{Cumulative weight fraction $p_K$ of the $K$ largest-weight determinants of the exact Fe$_2$S$_2$ [30e,\,20o] FCI ground state in the canonical Hartree--Fock (blue) and energy-optimized (red) orbital bases. By Eq.~\eqref{eq:truncated_ci_trial}, $p_K$ is the squared overlap of the normalized $K$-determinant truncation with the ground state, and because the weights are exact, each curve gives the maximal overlap attainable with $K$ determinants in that basis. Each marker gives $K_f$ (Eq.~\eqref{eq:fidelity-support}) in each basis, and arrows give the ratio $K_f^{\mathrm{HF}}/K_f^{\mathrm{opt}}$. 
    }
    \label{fig:fidelity-support}
\end{figure}

Beyond classical energy and cost benchmarks, SCI calculations provide explicit CI wavefunctions that can serve as trial states for QPE and inform its resource requirements~\cite{Fomichev_2024,tubman2018postponingorthogonalitycatastropheefficient}. Ground-state overlap governs the repetition overhead, while determinant support influences preparation cost. 

A general CI wavefunction can be written as
\begin{equation}
\begin{aligned}
    |\Psi\rangle &=
\sum_L \sum_{i_L\in I_L}
c_{i_L}|\Phi_{i_L}\rangle\\
&= c_0|\Phi_0\rangle +
\sum_{i_1\in I_1} c_{i_1} |\Phi_{i_1}\rangle +
\sum_{i_2\in I_2} c_{i_2} |\Phi_{i_2}\rangle + \cdots,
\end{aligned}
\end{equation}
where $L$ is the excitation level relative to the determinant used to define excitation rank, and $I_L$ denotes a set of determinant indices corresponding to $L$. $|\Phi_{i_1}\rangle$, $|\Phi_{i_2}\rangle$, $\ldots$ denote singly, doubly,
and higher excited determinants. For the Hartree--Fock basis, $|\Phi_0\rangle=|\text{HF}\rangle$ and $c_0 = c_{\text{HF}}$. Weights for each excitation level are given by
\begin{equation}
W_L
=
\sum_{i_L\in I_L}|c_{i_L}|^2,
\end{equation}
which should be interpreted as a basis-dependent weight decomposition of the wavefunction across excitation sectors.
In Fig.~\ref{fig:excited_lev}, Lv0 represents the weight of the zero-excitation determinant in this basis, $W_0 = |c_{\mathrm{ref}}|^2$.

We analyze excitation-level truncations of the computed CI states and optimal top-$K$ truncations of the exact Fe$_2$S$_2$ ground state to quantify how orbital choice affects the determinant support needed for a specified overlap. In canonical Hartree--Fock orbitals, the reference determinant carries approximately $82$--$90\%$ of the weight in the Ru systems, while configurations up to quadruple excitation account for $\sim60\%$ of the Fe$_2$S$_2$ wavefunction weight in both orbital representations. Capturing substantial CI overlap therefore requires different excitation support across these systems. 

A simple way to build a trial state from the reported CI state $|\Psi_{\mathrm{CI}}\rangle$ is to keep only determinants up to a fixed excitation level $L^{*}$, and the level weights $W_L$ of Fig.~\ref{fig:excited_lev} bound the overlap that any such state can reach. Let $\mathcal{S}_{\le L^{*}}$ denote the subspace spanned by all determinants through excitation level $L^{*}$. Then,
\begin{equation}
\max_{|\Phi\rangle\in\mathcal{S}_{\le L^{*}}}
|\langle\Psi|\Phi\rangle|^{2}
=
\sum_{L\le L^{*}} W_L ,
\label{eq:max-overlap}
\end{equation}
so the cumulative excitation-level weight gives the maximum target-state overlap attainable at a given excitation cutoff. Since the expected QPE repetition count scales inversely with the squared initial-state overlap, these weights directly inform the quantum cost~\cite{tubman2018postponingorthogonalitycatastropheefficient, Fomichev_2024,gratsea2025achievingutilityscaleapplicationsstack}.

Orbital optimization changes this assessment in several cases.  
In Fig.~\ref{fig:excited_lev}, the weight of the reference occupation pattern in FeMoco decreases from $44.7\%$ in canonical Hartree--Fock orbitals to $4.7\%$ in the energy-optimized basis, accompanied by a substantial redistribution of weight across excitation ranks. In contrast, the Ru systems exhibit smaller changes and the Fe$_2$S$_2$ profile remains nearly unchanged. We quantify this basis sensitivity using $L_1=\sum_L|W_L^{\mathrm{HF}}-W_L^{\mathrm{OPT}}|$: a larger value indicates that the excitation profile in Hartree--Fock orbitals does not carry over unchanged to the optimized representation. A small $L_1$ indicates that the canonical HF basis yields an excitation-level weight distribution similar to that obtained with energy-optimized orbitals.

However, excitation-level weights $W_L$ do not resolve the distribution of weight among determinants within each level, so similar excitation profiles can conceal substantial differences in the determinant support required for a given overlap. To characterize this determinant-level structure, we consider a trial state formed from the $K$ determinants with the largest coefficient magnitudes, irrespective of excitation level. Ordering the coefficients of $|\Psi_{\mathrm{CI}}\rangle$ by
$|c_{(1)}|^2\geq|c_{(2)}|^2\geq\cdots$ gives
\begin{equation}
p_K=\sum_{i=1}^{K}|c_{(i)}|^2,
\qquad
|\Phi_K\rangle=
\frac{1}{\sqrt{p_K}}
\sum_{i=1}^{K}c_{(i)}|D_{(i)}\rangle.
\label{eq:truncated_ci_trial}
\end{equation}
This construction preserves the relative coefficient phases and achieves
$|\langle\Psi_{\mathrm{CI}}|\Phi_K\rangle|^2=p_K$. Within the chosen
determinant basis, it maximizes the overlap among normalized expansions
containing at most $K$ determinants; in particular, for the same number of
determinants it is never worse than the excitation-rank truncation of
Eq.~\eqref{eq:max-overlap}. The minimum number of
determinants required to reach a target squared overlap $f$ is   
\begin{equation}
K_f=\min\{K:\,p_K\ge f\}.
\label{eq:fidelity-support}
\end{equation}
When $|\Psi_{\mathrm{CI}}\rangle$ is a variational selected-CI state, $p_K$ and $K_f$ refer to overlap with that approximate state, and overlap with the exact ground state additionally depends on the fidelity of $|\Psi_{\mathrm{CI}}\rangle$ itself. When $|\Psi_{\mathrm{CI}}\rangle$ is the exact FCI ground state, $K_f$ is the minimum support required to reach overlap $f$ with the ground state in that basis. 

Fig.~\ref{fig:fidelity-support} shows the minimum determinant support required to reach overlap $f$ in each orbital basis for Fe$_2$S$_2$, for which the exact FCI ground state is performed. Orbital optimization reduces this support by more than two orders of magnitude at moderate overlap, but the benefit of orbital optimization diminishes as the target overlap increases. The ratio $K_f^{\mathrm{HF}}/K_f^{\mathrm{opt}}$ falls from ${\sim}229$-fold at $f=0.5$ to ${\sim}30$-fold at $f=0.999$. Orbital optimization compresses the leading part of the ground state far more efficiently than its tail. 

Convergence of the variational energy alone does not establish convergence to the intended ground state, especially when multiple low-energy solutions compete, as illustrated by the metastable DMRG solutions in P450 (Appendix~\ref{app:p450_state_validation}). Overlaps with approximate reference states should therefore be distinguished from the actual ground-state success probabilities of QPE. For an exact reference, as in our Fe$_2$S$_2$ analysis, $p_K$ directly gives this probability in ideal QPE.

\section{Discussion}
\label{sec:discussion}

The value of quantum computing in chemistry depends on whether it can solve important chemical problems at a speed and cost that justify its use~\cite{10.1021/acs.jpclett.6c00888,Herrmann:2023sfw}. Our results show that classical computation remains cheaper for six of our seven benchmarks, even assuming fault-tolerant quantum hardware with sufficient logical qubits is available at US\$7{,}000 per whole-QPU hour. Thus, hardware availability alone does not guarantee a cost advantage. These results refine forecasts of quantum advantage based on asymptotic algorithmic scaling~\cite{Gundlach:2025sir} by highlighting practical challenges in chemistry. In addition to the advances in classical electronic-structure methods~\cite{zhai2026classicalcomputationalsimulationfemocofactor,zhang2026absorbingmanybodycorrelationscoreoptimized,shang2026closing}, these findings motivate a more selective search for quantum applications in chemistry, including problems beyond ground-state energy estimation, such as electron-dynamics simulation~\cite{Langkabel:2022djw}.

Nevertheless, the lack of an established generic exponential advantage~\cite{lee2023evaluating,chan2024spiers} does not rule out useful quantum speedups. Scientific urgency or otherwise inaccessible insights may justify such costs; COVID Moonshot, for example, illustrates how urgent needs can motivate substantial computational and experimental effort for a single therapeutic target~\cite{doi:10.1126/science.abo7201}. Shor's algorithm likewise illustrates the value of problem-specific quantum speedups, solving integer factorization in polynomial time, whereas no classical polynomial-time algorithm is yet known~\cite{Shor:1994jg}. Specialized, high-value applications may therefore justify continued development of quantum computing and its applications. 

Reliable classical references are essential to assessing whether quantum computing can deliver such value in chemistry. Our assessment of DMRG and SHCI references illustrates why apparently converged classical benchmarks require independent validation. In the P450 $2M_S=5$ sector, our variational energies lie up to $31.8$~mHa below the published DMRG references, with independent DMRG calculations reproducing the lower energies within $0.5$~mHa in active spaces $d$ and $e$. Calculations for $d$ further distinguish the capacity to represent a low-energy state from the ability to reach it through optimization: two $M=4000$ final runs satisfy convergence criteria despite their substantial energy separation, while the lower branch remains accessible at $M=512$. Energy stagnation therefore provides incomplete evidence of convergence, motivating tests of initialization and orbital ordering before attributing the remaining error to insufficient bond dimension.

Without such validation, difficulties encountered by a particular classical optimization procedure can be mistaken for intrinsic classical hardness. DMRG energies and MPSs are commonly used to assess Hamiltonian-compression errors, identify target symmetry sectors, and motivate high ground-state overlaps for quantum phase estimation (QPE)~\cite{DFTHC,Goings_2022,Fomichev_2024}. A metastable DMRG solution can therefore affect estimates of Hamiltonian-compression errors and lead to misleading estimates of trial-state success probabilities. An unreliable classical reference undermines claims that the encoded Hamiltonian preserves the intended low-energy physics, that the calculation achieves the stated accuracy, and that the target offers a meaningful quantum advantage. For systems without an exact reference, agreement within chemical accuracy between independently optimized methods under the same Hamiltonian and symmetry constraints provides a useful cross-check.

The trial-state analysis links classical wavefunction structure to
fault-tolerant QPE costs. For Fe$_2$S$_2$, orbital optimization reduces the determinant  support required for $99\%$ squared overlap by nearly 60-fold, despite largely unchanged excitation-level  weight distributions. This advantage diminishes as the overlap approaches unity, as optimization compresses dominant determinants more effectively than the wavefunction tail. The cumulative weights $p_K$ quantify the overlap available from a given determinant support, linking state-preparation requirements to QPE repetitions. Lower overlap increases quantum execution costs and reduces the break-even QPU price, reinforcing the classical cost advantage for six of the seven systems and particularly affecting FeMoco, whose break-even range spans the reference QPU price.

Recent advances in orbital optimization have shown that accurate wavefunctions can be represented using fewer determinants~\cite{zhang2026absorbingmanybodycorrelationscoreoptimized}. However, reducing the number of determinants does not necessarily reduce quantum state-preparation costs by the same factor. Our comparisons therefore distinguish representation size, computational work, and elapsed time, while treating transfers between solver-specific cost models as conditional estimates.

Our classical calculations were conducted on mid-range CPU clusters, Improv at Argonne National Laboratory~\cite{ANL_LCRC_2026,ANL_Improv_2026} and Spark at the University of Wisconsin--Madison, with extrapolation required for the most demanding systems. Larger computational resources could shorten time to solution, while algorithmic advances, as illustrated by recent progress on FeMoco~\cite{zhai2026classicalcomputationalsimulationfemocofactor,shang2026closing}, could further reduce computational work and cost.

\section*{Data Availability}
All data required to support the figures and tables
in this work are available on Zenodo~\cite{zenodo_23651694}.

\section*{Acknowledgements}
Calculations were performed on Spark at the University of Wisconsin-Madison and Improv, operated by the Laboratory Computing Resource Center at Argonne National Laboratory. This work was supported by the National Science Foundation through the Quantum Leap Challenge Institute for Hybrid Quantum Architectures and Networks (Award No. 2016136) and CAREER Grant No. 2543065, and by the U.S. Department of Energy, Office of Science, National Quantum Information Science Research Centers. We thank Andor Menczer for providing the raw DMRG data underlying the published reference energies~\cite{10.1021/acs.jctc.4c00903}, including the bond-dimension series in Table~\ref{tab:trimci_best}.
The authors used AI
tools to assist with manuscript drafting and editing, literature searches, and review of mathematical derivations
and resource estimates. The authors reviewed and revised the resulting material and take responsibility for
the final manuscript.

\bibliography{library}

\clearpage
\widetext

\begin{center}
\Large\textbf{APPENDICES}
\end{center}
\vspace{1em}

\noindent\textbf{Appendix \ref{sec:workflow}: Workflow}
\dotfill \pageref{sec:workflow} \\
\hspace*{1.5em} Fig.~\ref{fig:workflow}: Overview of the computational work
\dotfill \pageref{fig:workflow} \\
\hspace*{1.5em} Fig.~\ref{fig:four-routes}: Effect of orbital preparation
\dotfill \pageref{fig:four-routes} \\

\noindent\textbf{Appendix \ref{app:dmrg_support}: DMRG convergence and reference validation}
\dotfill \pageref{app:dmrg_support} \\
\hspace*{1.5em} Fig.~\ref{fig:dmrg-convergence}: DMRG convergence for P450 active spaces $d$ and $e$
\dotfill \pageref{fig:dmrg-convergence} \\

\noindent\textbf{Appendix \ref{app:p450_state_validation}: DMRG state identification and trial-state overlaps}
\dotfill \pageref{app:p450_state_validation} \\
\hspace*{1.5em} Table~\ref{tab:p450_overlap_validation_workload}: Trial-state overlaps and DMRG initialization and orbital ordering
\dotfill \pageref{tab:p450_overlap_validation_workload} \\

\noindent\textbf{Appendix \ref{app:ms-sector-ordering}: SHCI and TrimCI $M_S$-sector ordering}
\dotfill \pageref{app:ms-sector-ordering} \\
\hspace*{1.5em} Fig.~\ref{fig:shci_trimCI_combined}: Fixed-$M_S$ sector energies
\dotfill \pageref{fig:shci_trimCI_combined} \\

\noindent\textbf{Appendix \ref{app:pt2-sensitivity}: PT2 convergence-model sensitivity}
\dotfill \pageref{app:pt2-sensitivity} \\

\noindent\textbf{Appendix \ref{sec:femoco-estimation}: FeMoco projection protocol}
\dotfill \pageref{sec:femoco-estimation} \\
\hspace*{1.5em} Fig.~\ref{fig:femoco-projection}: FeMoco variational-energy projections
\dotfill \pageref{fig:femoco-projection} \\

\noindent\textbf{Appendix \ref{app:price}: Cost relation with cloud computing services}
\dotfill \pageref{app:price} \\
\hspace*{1.5em} Table~\ref{tab:illustrative_cloud_pricing}: Illustrative pricing inputs for the cost comparison
\dotfill \pageref{tab:illustrative_cloud_pricing} \\

\noindent\textbf{Appendix \ref{app:data-tables}: Benchmark data tables}
\dotfill \pageref{app:data-tables} \\
\hspace*{1.5em} Table~\ref{tab:trimci_best}: Best variational energies obtained with TrimCI
\dotfill \pageref{tab:trimci_best} \\
\hspace*{1.5em} Table~\ref{tab:allinone}: Summary of SHCI results and computational resources
\dotfill \pageref{tab:allinone} \\
\hspace*{1.5em} Table~\ref{tab:allinoneTrim}: Summary of TrimCI results and computational resources
\dotfill \pageref{tab:allinoneTrim} \\

\vspace{1em}

\appendix

\clearpage

\section{Workflow}
\label{sec:workflow}

\begin{figure}[htbp]
    \centering
    \includegraphics[width=1.0\linewidth]{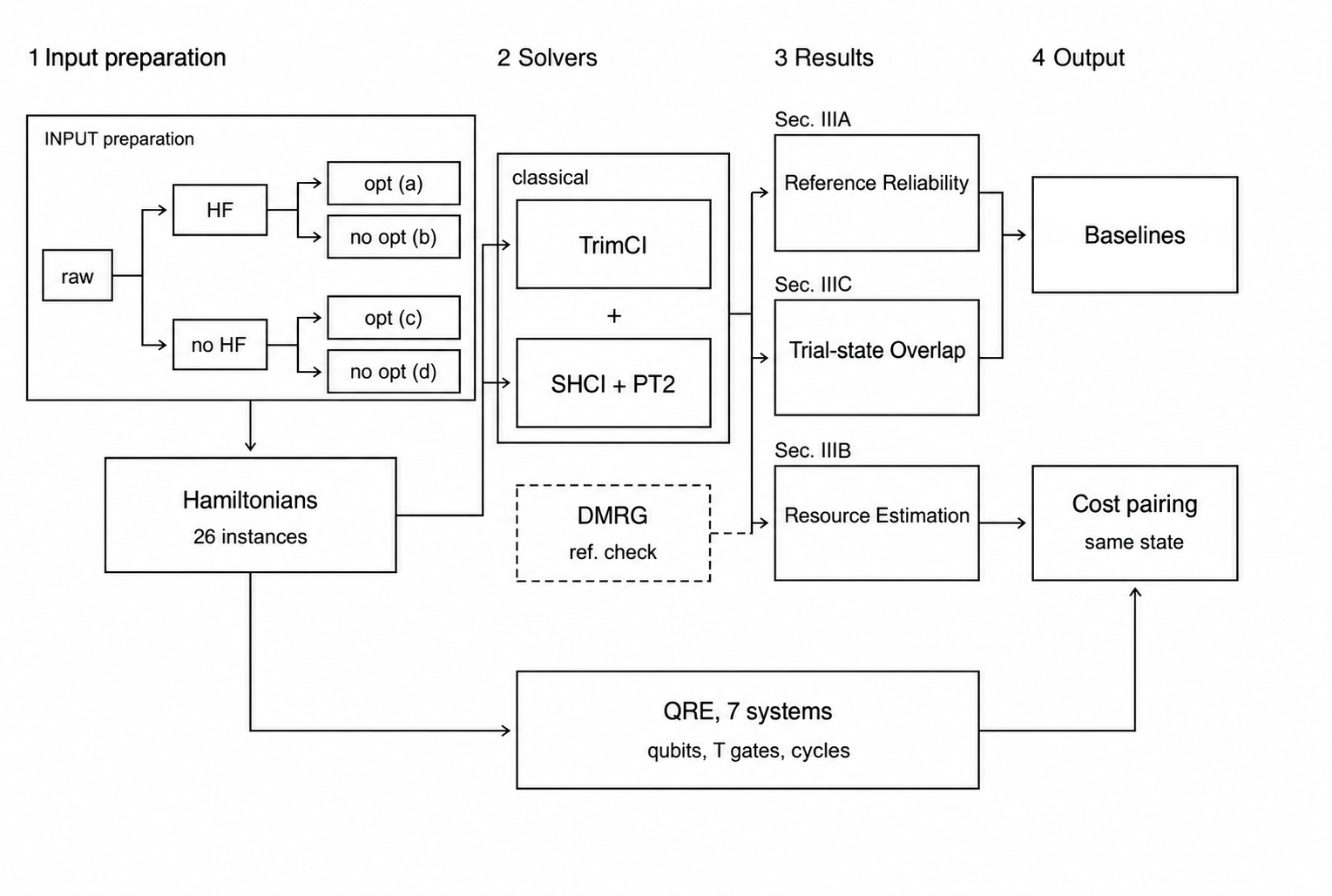}
        \caption{Overview of the computational work.}
    \label{fig:workflow}
\end{figure}
\begin{figure}[htbp]
    \centering
    \includegraphics[width=1.0\linewidth]{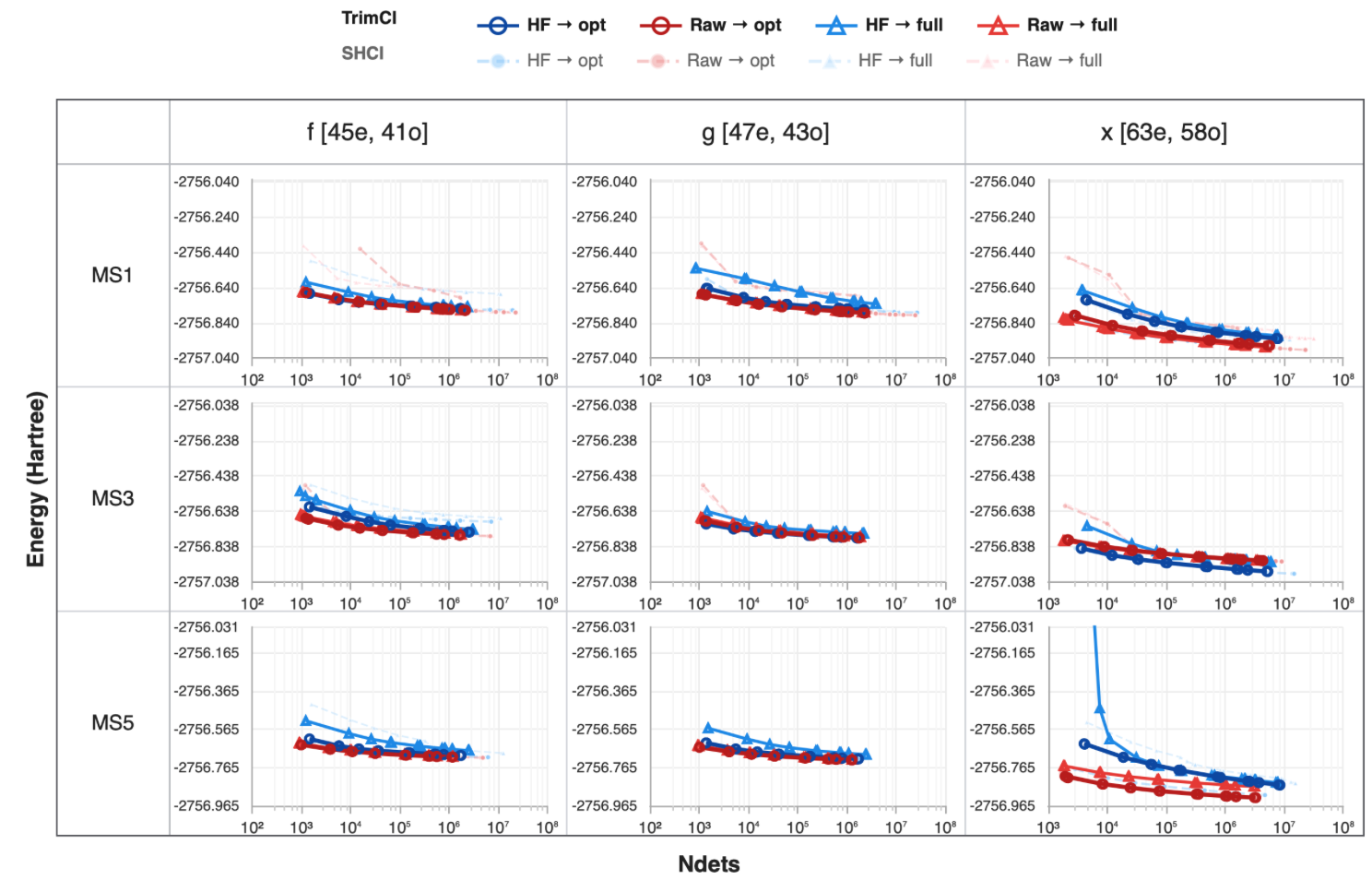}
    \caption{Effect of orbital preparation on variational energy convergence for P450 active spaces $f$ [45e, 41o], $g$ [47e, 43o], and $x$ [63e, 58o]. Columns correspond to active spaces and rows to spin sectors $2M_S=1,3,5$. Energies are plotted against the number of determinants, $N_{\mathrm{det}}$, for TrimCI (dark solid lines) and SHCI (light dashed lines). Blue and red indicate HF-based and raw-orbital starting points, respectively, while circles and triangles denote the OPT and Full preparation routes.}
    \label{fig:four-routes}
\end{figure}

We examined the consistency of the active-space orbitals through a Hartree--Fock (HF) convergence test and prepared the FCIDUMP inputs along four routes, branched by (i) whether an HF preparation step was applied and (ii) whether a subsequent orbital optimization (OPT) step was performed, yielding the four cases \textbf{(a)} HF \, $\rightarrow$ \, OPT, \textbf{(b)} HF without OPT, \textbf{(c)} raw \, $\rightarrow$ \, OPT, and \textbf{(d)} raw without OPT. We compare SHCI and TrimCI calculations with and without orbital optimization, following the respective procedures in~\cite{10.1021/acs.jctc.1c00385,zhang2025randomdeterminantsgroundstate}. These comparisons reveal how orbital choice affects convergence, wavefunction compactness, and excitation-level distributions. Starting from the raw active-space orbitals of~\cite{Goings_2022,Li_2018,LiFeMocoGitHub}, all four preparation routes were carried out before SHCI/TrimCI runs. For the majority of the SHCI calculations, we used the more standard workflow based on HF \, $\rightarrow$ \, OPT orbitals, i.e.\ case~(a).

For the P450 $f$, $g$, and $x$ active spaces, optimizing the raw orbitals provided by~\cite{Goings_2022} yielded negligible changes in SHCI variational energies (Fig.~\ref{fig:four-routes}), suggesting that these orbitals were already near optimal from the source~\cite{Goings_2022}.

Although TrimCI is more memory-efficient than SHCI (Tables~\ref{tab:allinone} and~\ref{tab:allinoneTrim}), we use SHCI timings for the cost-scaling analysis, because the cumulative work of TrimCI depends on its expansion and trimming history, not only on the final determinant count. On the other hand, tightening \(\epsilon_1\) in SHCI systematically expands the variational space and lowers the energy, thus providing an end-to-end measure of the classical cost of each SHCI final calculation point.

\clearpage

\section{DMRG Convergence and Reference Validation}
\label{app:dmrg_support}

\begin{figure*}[htbp]
    \centering
    \begin{subfigure}[t]{0.49\textwidth}
        \centering
        \includegraphics[width=\linewidth]{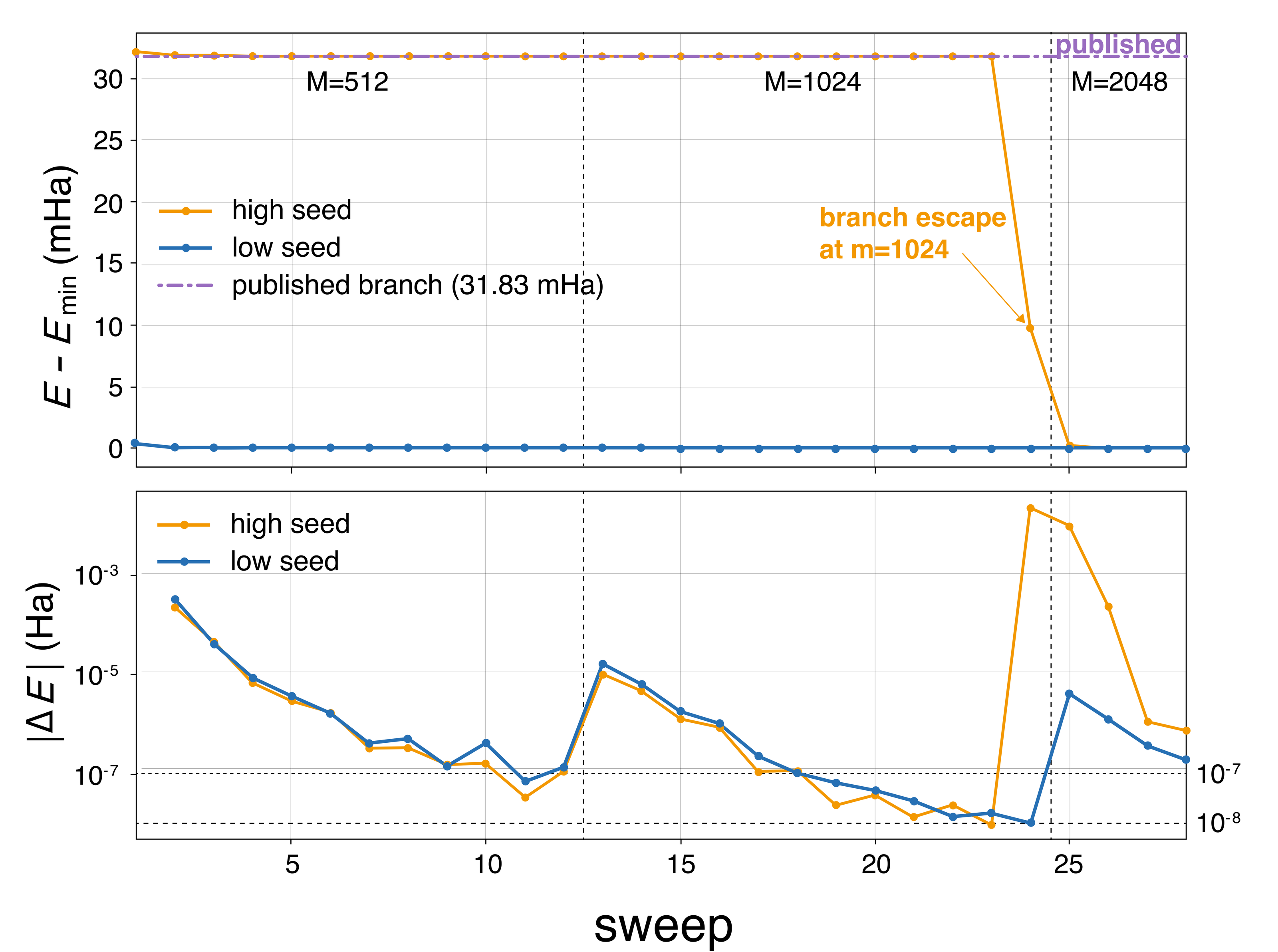}
        \caption{}
    \end{subfigure}
    \hfill
    \begin{subfigure}[t]{0.49\textwidth}
        \centering
        \includegraphics[width=\linewidth]{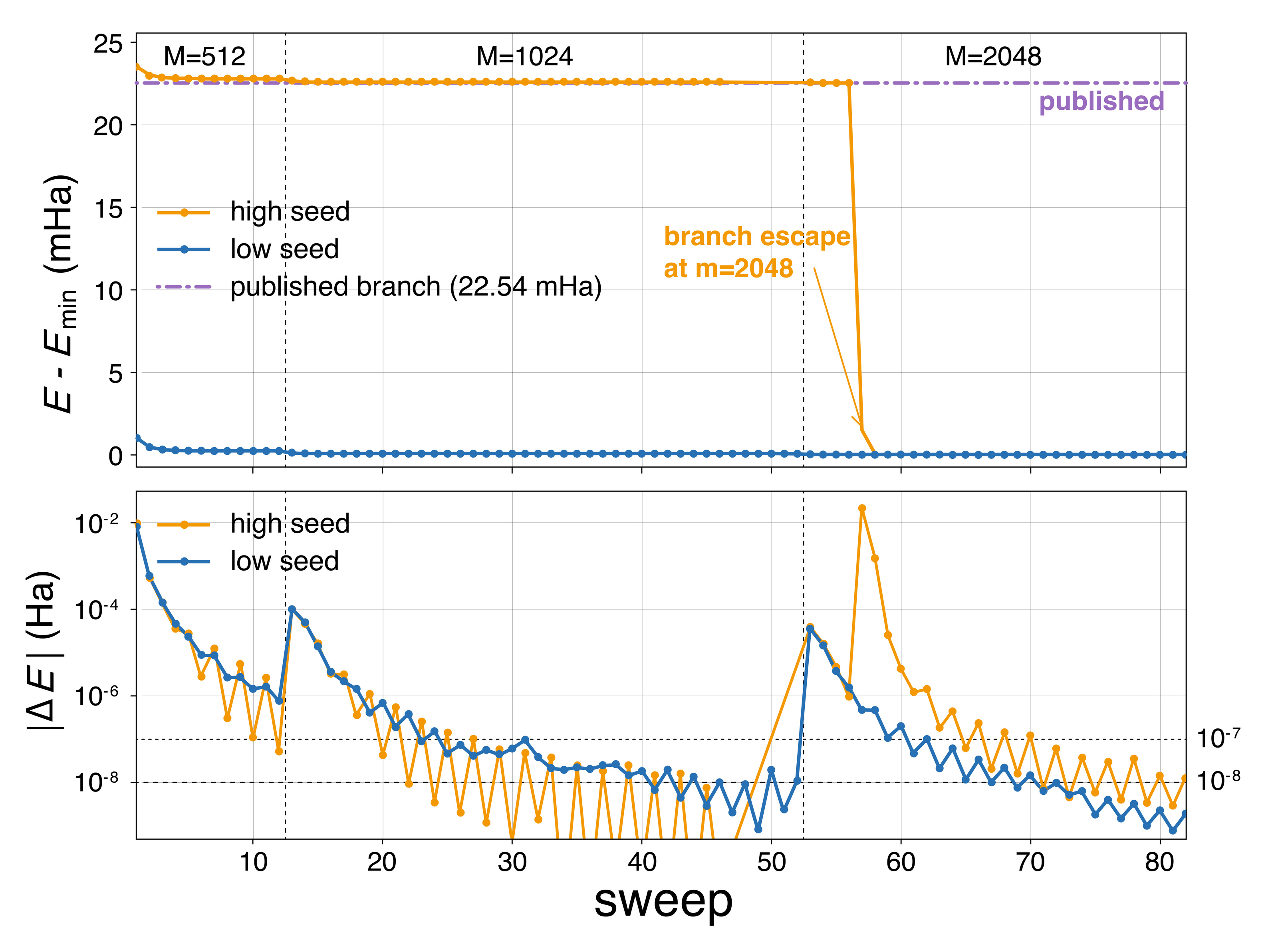}
        \caption{}
    \end{subfigure}
    \caption{
    DMRG convergence for P450 active spaces (a) $d$ [25e, 23o] and (b) $e$ [33e, 31o] in the $2M_S=5$ sector. In each space, two determinant-defined initial states are optimized using the same Fiedler orbital ordering and bond-dimension schedule, $M=512$, $1024$, and $2048$. Upper panels show energies relative to the lowest values encountered, $E_{\min}^{(d)}=-2756.539051099$~Ha and $E_{\min}^{(e)}=-2756.565094323$~Ha, respectively. Lower panels show absolute adjacent-sweep energy changes, $|\Delta E|=|E_s-E_{s-1}|$, with vertical dashed lines separating bond-dimension stages. (a) The low-energy seed reaches the lower branch at $M=512$, while the high-energy seed remains approximately $31.8$~mHa higher before a sharp decrease at $M=1024$ and subsequent approach to the lower branch at $M=2048$. The purple dash-dotted line marks the published reference~\cite{10.1021/acs.jctc.4c00903}. (b) The low-energy seed reaches the lower branch at $M=512$, whereas the high-energy seed remains approximately $22.6$~mHa higher through $M=1024$, despite an adjacent-sweep change of only $4.4\times10^{-11}$~Ha. At $M=2048$, its energy drops sharply at the fifth sweep, reducing the difference between trajectories to $0.735~\mu$Ha after 30 sweeps. Both cases demonstrate that small adjacent-sweep energy changes do not guarantee convergence to the lowest accessible variational state.
    }
    \label{fig:dmrg-convergence}
\end{figure*}

We performed DMRG calculations with block2 (v0.5.3)~\cite{10.1063/5.0180424}, using the non-spin-adapted, $S_z$-conserving formulation in the $2M_S=5$ sector. The trajectories in Fig.~\ref{fig:dmrg-convergence} use the same Fiedler orbital ordering~\cite{10.1063/1.4905329} and bond-dimension schedule, $M=512$, $1024$, and $2048$, starting from high- and low-energy states specified by determinant-space coefficient vectors.

Fig.~\ref{fig:dmrg-convergence}(a) shows distinct convergence trajectories in active space $d$. The low-energy seed reaches the lower branch at $M=512$, while the high-energy seed remains approximately $31.8$~mHa higher, near the published DMRG reference~\cite{10.1021/acs.jctc.4c00903}. The high-energy trajectory undergoes a sharp decrease at sweep~24 ($M=1024$) and approaches the lower branch.

Active space $e$ [33e, 31o] exhibits similar behavior (Fig.~\ref{fig:dmrg-convergence}(b)). The low-energy seed already reaches the lower branch at $M=512$, whereas the high-energy seed remains approximately $22.56$~mHa higher at its $M=1024$. At $M=2048$, the high-energy trajectory escapes this plateau, dropping by approximately $21.06$~mHa.

To assess eigenstate convergence independently of sweep-energy changes, we examine the residual $(H-E)|\Psi\rangle$, which vanishes for an exact eigenstate. For a higher-energy $M=1024$ checkpoint in active space $e$, the projected residual gives a lower bound of $15.88$~mHa on the Hamiltonian standard deviation, despite negligible changes in sweep energy.

Separately, we performed $M=4000$ calculations to test whether distinct final DMRG solutions persist under different initialization and orbital-ordering histories. For active space $d$, a cold start in the Fiedler ordering yielded a higher-energy solution, whereas initialization in the original ordering followed by re-optimization in the Fiedler ordering yielded a lower-energy solution:
\begin{align}
E_{d,\mathrm{Fiedler,high}}^{M=4000}
&= -2756.507220241~\mathrm{Ha},\\
E_{d,\mathrm{Fiedler,low}}^{M=4000}
&= -2756.539052246~\mathrm{Ha}.
\end{align}
The final adjacent-sweep energy changes were $-2.19\times10^{-9}$ and $-1.92\times10^{-9}$~Ha, respectively. Despite satisfying the convergence criterion, the solutions remained separated by $31.832$~mHa. The higher energy reproduces the published DMRG reference, while the lower agrees within $4.62~\mu\mathrm{Ha}$ with the TrimCI variational energy, $E_{d,\mathrm{var}}^{\mathrm{TrimCI}}=-2756.5390476269$~Ha (Table~\ref{tab:trimci_best}).

For active space $e$, separate cold $M=4000$ calculations in the Fiedler and original orderings yielded
\begin{align}
E_{e,\mathrm{Fiedler,high}}^{M=4000}
&= -2756.542556261~\mathrm{Ha},\\
E_{e,\mathrm{original,low}}^{M=4000}
&= -2756.564921993~\mathrm{Ha},
\end{align}
a difference of $22.366$~mHa. A similar energy separation arises from different initial states even within the same Fiedler ordering, as shown in Fig.~\ref{fig:dmrg-convergence}(b).

In both active spaces, the higher- and lower-energy solutions are consistent with nearly pure sextet states ($S=5/2$), with higher-spin contamination bounded $<3.1\times10^{-6}$. The energy separation thus arises within the same spin sector, and energy convergence alone cannot distinguish the two branches.

\clearpage

\section{DMRG state identification and trial-state overlaps}
\label{app:p450_state_validation}

To examine how DMRG convergence affects quantum trial-state preparation, we extracted sparse determinant expansions from the two DMRG branches in P450 active spaces~$d$ and $e$ and computed their squared overlaps with independently optimized SHCI wavefunctions. In both active spaces, trial states from the lower-energy branch (Appendix~\ref{app:dmrg_support}) have squared overlaps above $0.99$, while those from the higher-energy branch have overlaps below $10^{-10}$, even when the same number of determinants is retained (Table~\ref{tab:p450_overlap_validation_workload}(a)). The energy difference of $31.83$~mHa in active space~$d$ agrees with the first-excitation gap reported in~\cite{Schroeder:2026zkz}, suggesting that the higher-energy branch corresponds closely to the first excited state in the same spin sector.

For active space~$d$, we also examined the sensitivity of DMRG convergence to orbital ordering and MPS initialization (Table~\ref{tab:p450_overlap_validation_workload}(b)). DMRG with Fiedler ordering and unguided random-MPS initialization converged to the higher-energy branch for all five random seeds at the scheduled stopping point. Changing the orbital ordering, introducing ROHF occupation guidance, or both recovered the lower-energy solution in all 15 remaining runs~\cite{10.1063/1.4905329,10.1063/5.0180424}.

\begin{table*}[htbp]
\centering
\caption{
Trial-state overlaps and sensitivity of DMRG convergence to orbital ordering and initialization.
\textbf{(a)} Squared overlaps $p=|\langle\Psi_{\mathrm{trial}}|\Psi_{\mathrm{SHCI}}\rangle|^2$ with independently optimized SHCI reference states. We retain the $K$ determinants with the largest coefficient magnitudes in each branch, where $K$ is the number of determinants with $|c_I|\geq10^{-3}$ in the lower-energy branch. Using the same coefficient cutoff in both branches leads to the same conclusion.
\textbf{(b)} Four combinations of orbital ordering and MPS initialization tested for active space~$d$, using the same five random seeds for each. The last column reports the number of runs that reached the lower-energy branch at the scheduled stopping point at $M=4000$. Random MPSs were generated using block2~\cite{10.1063/5.0180424}, and the Fiedler ordering follows~\cite{10.1063/1.4905329}.
}
\label{tab:p450_overlap_validation_workload}
\vspace{0.6em}
\footnotesize

\begin{minipage}[t]{0.35\textwidth}
\centering
\textbf{(a)} Trial-state overlap with SHCI\par\medskip
\begin{adjustbox}{max width=\linewidth}
\setlength{\tabcolsep}{4pt}
\begin{tabular}{llrr}
\toprule
Space & Branch & $K$ & $p$ \\
\midrule
$d$ & Lower  & 1{,}074 & 0.998735 \\
    & Higher & 1{,}074 & $8.66\times10^{-12}$ \\
\addlinespace
$e$ & Lower  & 2{,}764 & 0.995806 \\
    & Higher & 2{,}764 & $1.43\times10^{-12}$ \\
\bottomrule
\end{tabular}
\end{adjustbox}
\end{minipage}%
\hfill
\begin{minipage}[t]{0.62\textwidth}
\centering
\textbf{(b)} DMRG initialization and orbital ordering for P450-$d$\par\medskip
\begin{adjustbox}{max width=\linewidth}
\setlength{\tabcolsep}{4pt}
\begin{tabular}{cllc}
\toprule
Path & Orbital ordering & Random-MPS initialization & Lower branch \\
\midrule
P1 & Fiedler  & Unguided (\texttt{occs=None}) & 0/5 \\
P2 & Fiedler  & ROHF-guided ($\texttt{occs}=\mathbf{n}^{\mathrm{ROHF}}$) & 5/5 \\
P3 & Original & Unguided (\texttt{occs=None}) & 5/5 \\
P4 & Original & ROHF-guided ($\texttt{occs}=\mathbf{n}^{\mathrm{ROHF}}$) & 5/5 \\
\bottomrule
\end{tabular}
\end{adjustbox}
\end{minipage}
\end{table*}

\clearpage
\section{SHCI and TrimCI $M_S$-sector ordering}
\label{app:ms-sector-ordering}

\begin{figure*}[htbp]
    \centering
    \includegraphics[width=0.9\linewidth]{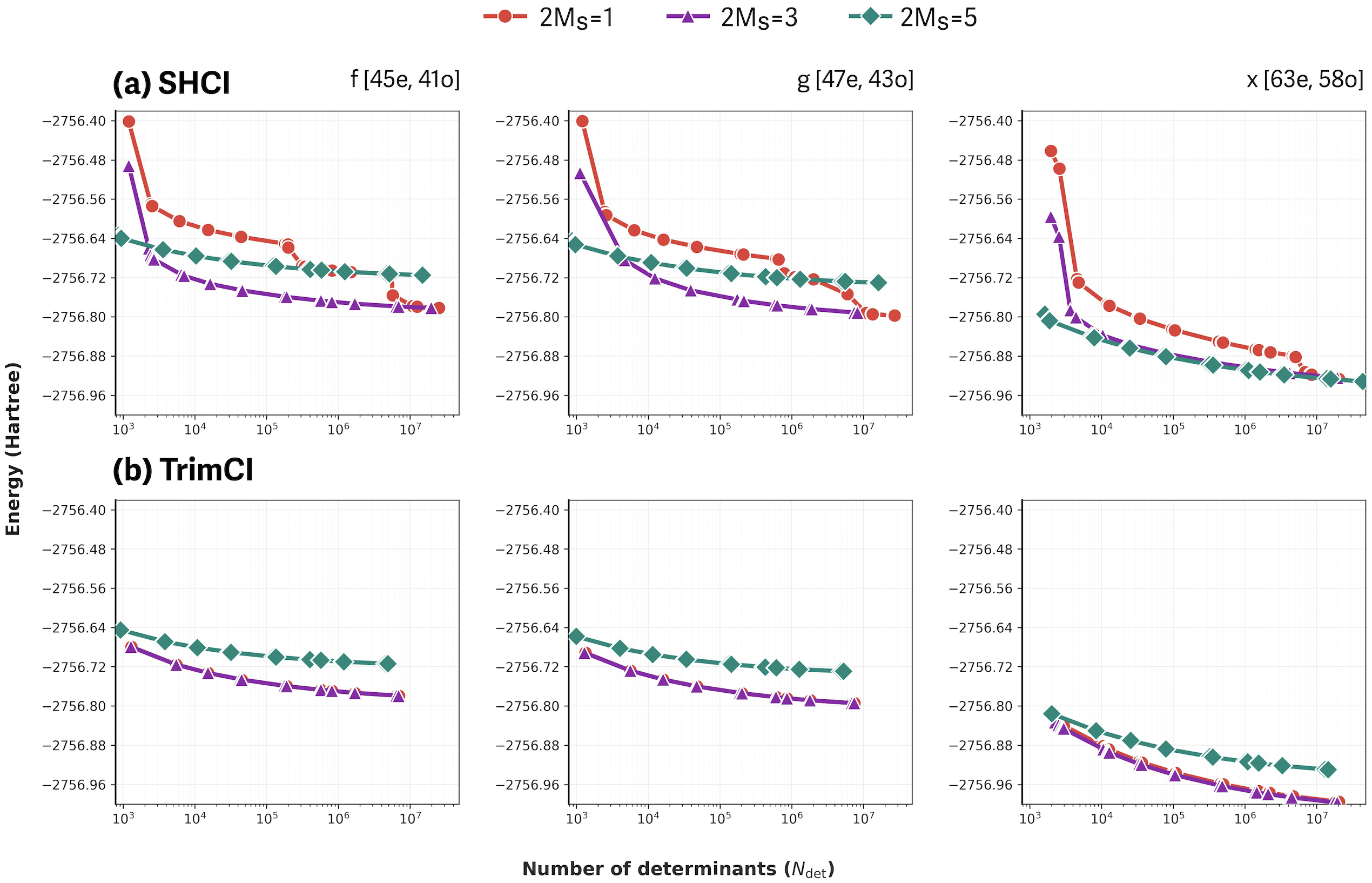}
    \caption{
    Comparison of fixed-$M_S$ sector energies versus variational-core size for (a) single-reference SHCI initialized from the Hartree--Fock determinant and (b) TrimCI-seeded SHCI in P450 Cpd~I active spaces $f$, $g$, and $x$. Single-reference SHCI shows strong finite-space sensitivity and can spuriously reverse the ordering of the $2M_S=1$, $3$, and $5$ sectors at small $N_{\mathrm{det}}$. In contrast, TrimCI-seeded SHCI maintains stable sector ordering across the expansion.
    }
    \label{fig:shci_trimCI_combined}
\end{figure*}

The SHCI selection criterion in Eq.~\eqref{eq:shci_selection} depends explicitly on the current coefficients $c_i$ and is therefore state- and reference- dependent~\cite{doi:10.1021/acs.jctc.6b00407,Sharma_2017}. Because the selection functional $|H_{ai}c_i|$ in eq.~\ref{eq:shci_selection} promotes configurations relative to the current variational state, sectors dominated by different determinant structures can be treated unevenly. This imbalance is especially important when comparing near-degenerate states across fixed-$M_S=S_z$ sectors. Under aggressive truncation, the resulting differential variational errors can exceed the true spin gap and spuriously invert the apparent ordering.

This behavior appears in Fig.~\ref{fig:shci_trimCI_combined}, where the SHCI energies of the $2M_S=1$, $2M_S=3$, and $2M_S=5$ sectors cross as a function of $N_{\mathrm{det}}$. Concretely, at $N_{\mathrm{det}} \approx 1\times10^{3}$ in the $[63e,\,58o]$ active space, SHCI places the $2M_S=5$ sector below $2M_S=3$ by ${\sim}200$~mHa. Only beyond $N_{\mathrm{det}} \approx 10^{7}$ does the ordering invert to the physically expected arrangement, and even there the lowest two sectors fluctuate by several mHa. 

By contrast, TrimCI maintains a consistent sector ordering over all $N_{\mathrm{det}}$ examined: the $2M_S=1$ and $2M_S=3$ sectors remain nearly degenerate, while $2M_S=5$ stays higher without ordering reversals. This behavior is consistent with experiment and other multireference calculations, which place the doublet and quartet close in energy and the sextet well above them~\cite{doi:10.1021/ja0424732,Goings_2022}.

The reversal of $M_S$-sector ordering under aggressive SHCI truncation shows that finite selected spaces can fail qualitatively in near-degenerate multireference regimes, where reference-dependent selection treats competing states unevenly. In the present calculations, TrimCI's stable sector ordering and tighter variational bounds are consistent with a more balanced treatment of the competing states.

\clearpage

\section{PT2 convergence-model sensitivity}
\label{app:pt2-sensitivity}

We first estimate the number of determinants $N^\star$ needed to reach $|E_{\mathrm{PT2}}|=1.6$~mHa. When the PT2 data include points on both sides of the threshold, we interpolate between the two neighboring points on a log--log scale. Otherwise, we extrapolate using a power-law fit to the last five data points.

To further examine how sensitive our estimates are to the choice of fitting function and fitting range, we use two models and fit each to the last $k\in\{4,5,6\}$ available PT2 data points:
\begin{align}
y &= a+bL, &&\text{power law},\\
y &= a+bL+cL^2, &&\text{log-quadratic},
\end{align}
where $L=\ln(N/N_{\max}^{(E)})$, $y=\ln(|E_{\mathrm{PT2}}|/1\,\mathrm{mHa})$, and $N_{\max}^{(E)}$ is the largest determinant count for which a PT2 calculation was completed. We use ordinary least squares, giving equal weight to all data points and allowing the intercept to vary.

For each fit, we determine $L^\star$ at $y=\ln(1.6)$ and obtain the corresponding determinant count, $N^\star=N_{\max}^{(E)}\exp(L^\star)$. We require the fitted PT2 magnitude to decrease with increasing $N$, both within the fitting range and throughout any extrapolation to $N^\star$. For log-quadratic extrapolations, we additionally require $c\leq0$ to prevent the convergence rate from slowing in log--log space. All 40 fits satisfy these requirements.

We then estimate the allocated CPU work needed to reach each $N^\star$ using the same molecule-specific model,
\begin{equation}
T^\star=A\left(\frac{N^\star}{N_0}\right)^\beta,
\end{equation}
while keeping the QPU runtimes and pricing assumptions fixed. The ranges shown in Fig.~\ref{fig:quantum-gain}(a--d) extend from the smallest to the largest estimates obtained from these fits.

For Fe$_4$S$_4$, all fitted determinant counts, $N^\star=(4.81\text{--}4.92)\times10^8$, fall within the available PT2 data range. However, complete CPU timing data are available only up to $4.24\times10^8$ determinants because the final PT2 data point lacks a complete record of the variational calculation times. Therefore, although no extrapolation of the PT2 data is needed, the CPU work model must be extended to determinant counts approximately 13--16\% beyond the timing data.

\clearpage
\section{FeMoco projection protocol}
\label{sec:femoco-estimation}

\begin{figure}[htbp]
    \centering
    \includegraphics[width=1\linewidth]{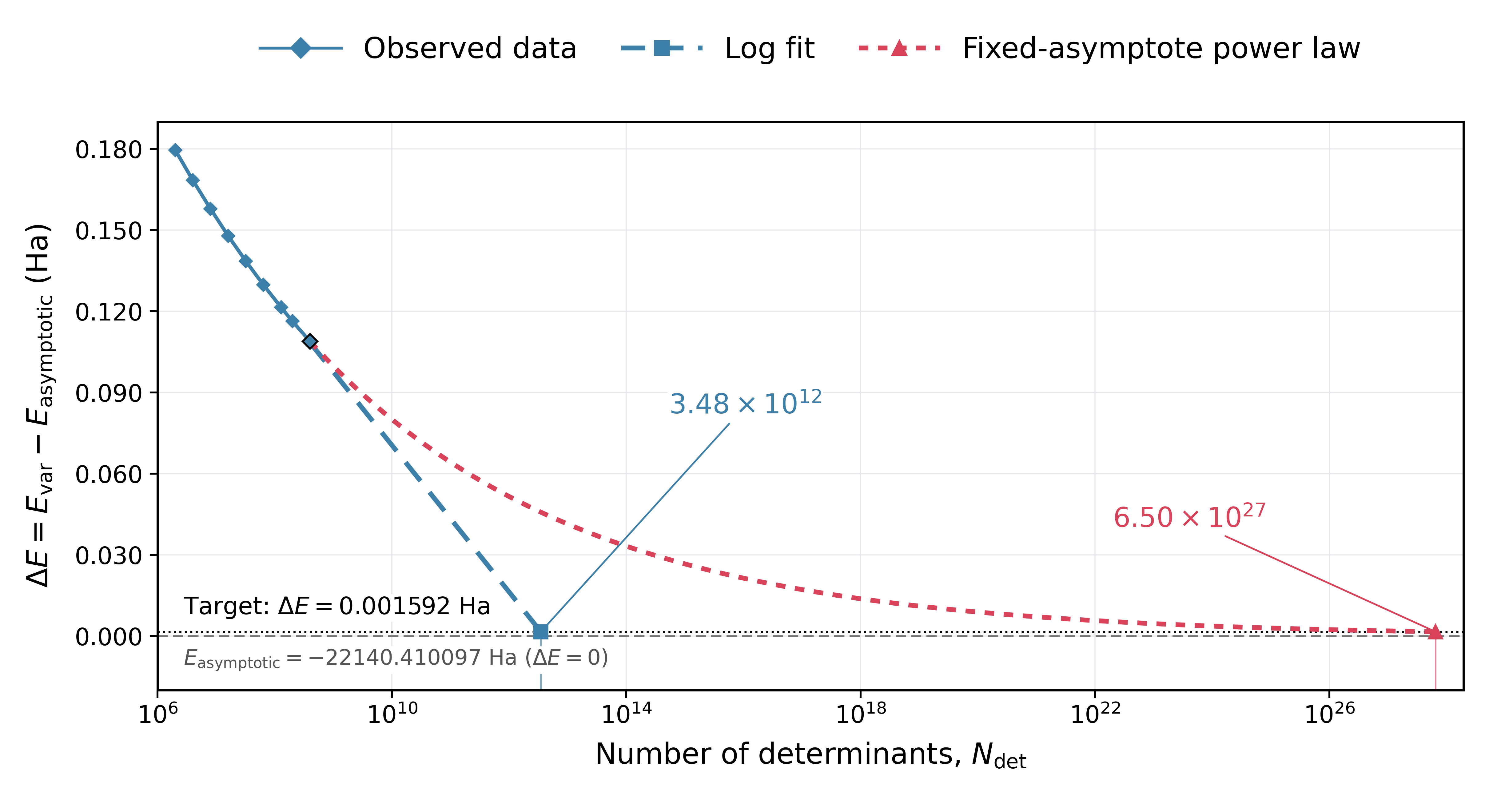}
    \caption{TrimCI variational error for the FeMo-cofactor LLDUC model [113e, 76o], measured against the infinite-bond-dimension UDMRG reference $E_{\mathrm{ref}}=-22140.410097$~Ha of~\cite{zhai2026classicalcomputationalsimulationfemocofactor}.  Logarithmic and fixed-asymptote power-law fits to the five largest of nine calculated points reach the $0.001592$~Ha target ($\approx1$~kcal/mol) at $3.48\times10^{12}$ and $6.50\times10^{27}$ determinants. We include the two extreme variational-energy extrapolations as a model sensitivity analysis.
    }
    \label{fig:femoco-projection}
\end{figure}

Fig.~\ref{fig:femoco-projection} estimates the number of determinants required for the TrimCI variational energy $E_{\mathrm{var}}$ to reach the literature criterion for chemical accuracy, $E_{\mathrm{ref}}=-22140.410097$~Ha and $E_{\mathrm{target}}=-22140.408505$~Ha from~\cite{zhai2026classicalcomputationalsimulationfemocofactor}; their energy difference of $0.001592$~Ha ($\sim1$~kcal/mol) is essentially equal to the chemical-accuracy threshold.

The figure shows TrimCI data points obtained for $N_{\mathrm{det}}=2\times10^{6}$--$4\times10^{8}$ and corresponding $\Delta E=E_{\mathrm{var}}^{\mathrm{TrimCI}}-E_{\mathrm{var}}^{\mathrm{ref}}$. To the five points with the largest determinant counts, we fitted a logarithmic model $\Delta E=a+b\log_{10}(N_{\mathrm{det}}/N_{0})$ and a power-law model $\Delta E=A\,(N_{\mathrm{det}}/N_{0})^{-p}$ whose asymptote is fixed at the reference energy, in both cases by unweighted least squares on the energy residuals with $N_{0}=4\times10^{8}$.  The logarithmic and fixed-asymptote power-law models reach the target at $3.48\times10^{12}$ and $6.50\times10^{27}$ determinants, respectively. The power-law extrapolation extends approximately nineteen orders of magnitude beyond our largest calculation, highlighting the dependence of the predicted requirement on the assumed convergence form.

The determinant targets derived from TrimCI variational energies are converted to allocated CPU work using the SHCI timing model in Fig.~\ref{fig:quantum-gain}(b). Because cumulative TrimCI work depends on its expansion and trimming history as well as the final determinant count, this conversion assumes that the SHCI work--determinant relation is transferable to TrimCI. The resulting cost estimates are therefore conditional on this assumption.

\clearpage
\section{Cost relation with cloud computing services}
\label{app:price}

\begin{table*}[htbp]
\centering
\small
\setlength{\tabcolsep}{8pt}
\renewcommand{\arraystretch}{1.15}
\caption{Illustrative pricing inputs for the cost comparison.
The CPU rate uses the August 30, 2026 Linux On-Demand quote
for US East (Ohio).
The QPU rate is based on dedicated IonQ Forte access through
Amazon Braket Direct; applying this rate to the modeled
fault-tolerant QPU is an illustrative assumption
\cite{AWSEC2OnDemandPricing2026,AWSHpc8a2026,AWSBraketPricing2026}.}
\label{tab:illustrative_cloud_pricing}

\begin{tabular}{@{}lll@{}}
\toprule
Service & Reference price & Rate used in this work \\
\midrule
EC2 \texttt{hpc8a.96xlarge}
& \$7.92 per instance-hour
& \$0.04125 per core-hour (192 physical cores) \\
Braket Direct, IonQ Forte
& \$7{,}000 per machine-hour
& \$7{,}000 per whole-QPU hour (exclusive access) \\
\bottomrule
\end{tabular}
\end{table*}

\clearpage

\section{Benchmark Data Tables}
\label{app:data-tables}

\begin{table*}[htbp]
\centering
\renewcommand{\arraystretch}{1.36}
\setlength{\tabcolsep}{6pt}
\LARGE
\resizebox{\textwidth}{!}{%
\begin{tabular}{|c|c|c|c|c|c|c|c|c|c|c|c|c|c|}
\hline
\multicolumn{1}{|c|}{System} &
\multicolumn{1}{c|}{Spin} &
\multicolumn{4}{c|}{System Size} &
\multicolumn{3}{c|}{Ground-State Energy Estimation} &
\multicolumn{3}{c|}{Reference Data\textsuperscript{\cite{Goings_2022,10.1021/acs.jctc.4c00903,doi:10.1021/acs.jctc.7b00270,zhai2026classicalcomputationalsimulationfemocofactor}}} &
\multicolumn{2}{c|}{Energy Difference (Ha)} \\
\hline
Molecule &
$2M_S$ &
\makecell{Space\\Definition} &
\makecell{$[N_e, N_o]$} &
\makecell{$N_{\mathrm{det}}$} &
Cluster &
$\epsilon_{\mathrm{var}}$ &
\makecell{$E_{\mathrm{var}}$\\(Ha)} &
\makecell{$E_{\mathrm{FCI}}$\\(Ha)} &
\makecell{DMRG\\$E_{\mathrm{var}}$ (Ha)} &
\makecell{DMRG\\extrapolated\\to $M\to\infty$ (Ha)} &
\makecell{CCSD(T)\\(Ha)} &
\makecell{$\Delta E_{\mathrm{var}}$ vs.\\ DMRG$_{M\to\infty}$} &
\makecell{$\Delta E_{\mathrm{var}}$ vs.\\ CCSD(T)} \\
\hline
\multirow{18}{*}{P450} &
\multirow{6}{*}{1} & c & [17e, 15o] & 4{,}164{,}969     & Spark & 5.00E-09 & -2756.5570756723 & $-2756.5570756690$ & -2756.55708 & -2756.557080 & -2756.556815 & $+0.000004$ & $-0.000261$ \\
& & d & [25e, 23o] & 58{,}660{,}978    & Spark & 1.00E-07 & -2756.5857121209 & -- & -2756.58571 & -2756.585720 & -2756.582548 & $+0.000008$ & $-0.003164$ \\
& & e & [33e, 31o] & 120{,}742{,}842   & Spark & 5.00E-07 & -2756.6210381859 & -- & -2756.62106 & -2756.621080 & -2756.619757 & $+0.000042$ & $-0.001281$ \\
& & f & [45e, 41o] & 1{,}045{,}498{,}058 & ANL & 5.00E-07 & -2756.7876003900 & -- & -2756.78809 & -2756.788480 & -2756.780654 & $+0.000880$ & $-0.006946$ \\
& & g & [47e, 43o] & 456{,}874{,}076   & ANL   & 1.25E-06 & -2756.8023460759 & -- & -2756.80362 & -2756.804620 & -2756.796661 & $+0.002274$ & $-0.005685$ \\
& & x & [63e, 58o] & 233{,}102{,}706   & ANL   & 4.50E-06 & -2757.0064659412 & -- & -2757.01629 & -2757.018870 & -2757.010201 & $+0.012404$ & $+0.003735$ \\
\cline{2-14}
& \multirow{6}{*}{3} & c & [17e, 15o] & 76{,}494          & Spark & 2.00E-06 & -2756.5569995644 & $-2756.5569996652$ & -- & -2756.556990 & -2756.556924 & $-0.000010$ & $-0.000076$ \\
& & d & [25e, 23o] & 2{,}689{,}396     & Spark & 2.00E-06 & -2756.5856102627 & -- & -- & -2756.585600 & -2756.584781 & $-0.000010$ & $-0.000829$ \\
& & e & [33e, 31o] & 32{,}190{,}406    & Spark & 1.00E-06 & -2756.6209440645 & -- & -- & -2756.620620 & -2756.619692 & $-0.000324$ & $-0.001252$ \\
& & f & [45e, 41o] & 358{,}775{,}540   & ANL   & 1.20E-06 & -2756.7868271446 & -- & -- & -2756.788440 & -2756.783417 & $+0.001613$ & $-0.003410$ \\
& & g & [47e, 43o] & 349{,}795{,}893   & ANL   & 1.50E-06 & -2756.8020773642 & -- & -- & -2756.804510 & -2756.799439 & $+0.002433$ & $-0.002638$ \\
& & x & [63e, 58o] & 385{,}609{,}266   & ANL   & 3.00E-06 & -2757.0099376435 & -- & -- & -2757.018470 & -2757.013449 & $+0.008532$ & $+0.003511$ \\
\cline{2-14}
& \multirow{6}{*}{5} & c & [17e, 15o] & 35{,}517          & Spark & 2.00E-06 & -2756.5113612535 & $-2756.5113612985$ & -2756.51136 & -2756.511360 & -2756.511293 & $-0.000001$ & $-0.000068$ \\
& & d & [25e, 23o] & 1{,}736{,}326     & Spark & 2.00E-06 & -2756.5390476269 & -- & -2756.50722 & -2756.50722 & -2756.53824 & $-0.031828$ & $-0.000808$ \\
& & e & [33e, 31o] & 827{,}868         & Spark & 2.00E-05 & -2756.5646481128 & -- & -2756.54255 & -2756.54255 & -2756.563936 & $-0.022098$ & $-0.000712$ \\
& & f & [45e, 41o] & 39{,}146{,}590    & Spark & 5.00E-06 & -2756.7172959853 & -- & -2756.70946 & -2756.709929 & -2756.716121 & $-0.007367$ & $-0.001175$ \\
& & g\textsuperscript{$\dagger$} & [47e, 43o] & 42{,}684{,}565    & Spark & 5.00E-06 & -2756.7330534270 & -- & -2756.72497 & -2756.726003 & -2756.732128 & $-0.007050$ & $-0.000925$ \\
& & x & [63e, 58o] & 379{,}583{,}811   & ANL   & 2.50E-06 & -2756.9400069405 & -- & -2756.94343 & -2756.948520 & -2756.502796 & $+0.008513$ & -- \\
\hline
Fe$_2$S$_2$ & 0 & -- & [30e, 20o] & 49{,}443{,}908 & Spark & 2.50E-08 & $-116.6056091$ & $-116.6056091204$ & $-116.60561$ & -- & -- & $+0.000001$ & -- \\
\hline
\multirow{2}{*}{Fe$_4$S$_4$} & \multirow{2}{*}{0} & \multirow{2}{*}{--} & \multirow{2}{*}{[54e, 36o]} & 1{,}301{,}867{,}594 & ANL & 6.00E-07 & $-327.2397785369$ & \multirow{2}{*}{--} & \multirow{2}{*}{$-327.2396369$} & \multirow{2}{*}{$-327.244300$} & \multirow{2}{*}{--} & $+0.004521$ & -- \\
& & & & 5{,}120{,}000{,}000\textsuperscript{\cite{zhang2026absorbingmanybodycorrelationscoreoptimized}} & ANL & 3.50E-07 & $-327.242156$ & & & & & $+0.002144$ & -- \\
\hline
FeMoco\textsuperscript{$\ddagger$} & 3 & -- & [113e, 76o] & 400{,}000{,}000 & ANL & -- & $-22140.3012456515$ & -- & $-22140.396133$ & $-22140.410097$ & $-22140.350447$ & $+0.108851$ & $+0.049201$ \\
\hline
\end{tabular}
} 
\vspace{1pt}
\begin{minipage}{\textwidth}
\footnotesize\justifying
\noindent\textsuperscript{$\dagger$}For the $2M_S=5$ $g$ active space, an independent spin-adapted DMRG calculation with bond dimension $D=1500$ reported $E=-2756.735600$~Ha~\cite{PhysRevLett.133.250601}, which is $2.547$~mHa below the present TrimCI variational energy and $9.597$~mHa below the published $M\to\infty$ DMRG extrapolation. Accordingly, the $g$ result is not claimed here as a new literature-wide variational upper bound.\\ \textsuperscript{$\ddagger$}The 400-million-determinant result was obtained by expanding a saved 200-million-determinant wavefunction to a scheduled determinant count using an adaptively adjusted selection threshold; hence no single $\epsilon_{\mathrm{var}}$ applies. For the LLDUC FeMoco model with $[113e,76o]$ and $2M_S=3$, the three reference columns consistently use the broken-symmetry BS8-E (BS8-237) results of~\cite{zhai2026classicalcomputationalsimulationfemocofactor}. The UDMRG extrapolation is an estimated limiting energy, not an exact FCI value, so the FCI entry is left blank.
\end{minipage}
\caption{Best variational energies obtained with TrimCI. We report only the rigorous variational upper bound in TrimCI, excluding the PT2 correction and extrapolation here. Thus, $E^{\mathrm{FCI}}\le E_{\mathrm{var}}^{\mathrm{TrimCI}}$ holds irrespective of the convergence status of any reference calculation; the two rightmost columns report energy differences relative to the published DMRG and CCSD(T) reference values. Here $\Delta E = E_{\mathrm{var}}^{\mathrm{TrimCI}} - E^{\mathrm{ref}}$ in Ha, and a negative value means that the TrimCI variational energy alone, without PT2 correction, lies below the reference. Of the 18 P450 active-space and spin-sector combinations, TrimCI lies below the $M\to\infty$ DMRG extrapolation in the original reports in seven, in two distinct groups; the $2M_S=5$ $g$ case is excluded from this count for the reason given in footnote~$\dagger$. For $2M_S=5$ with the $d$, $e$, and $f$ active spaces it lies below both the DMRG extrapolation ($0.0074$--$0.0318$~Ha) and the independently reported CCSD(T) energy ($0.0007$--$0.0012$~Ha). Two independent calculations locate variational energies
$31.8$ and $22.1$~mHa below the published values for $d$ and $e$. The other four ($2M_S=3$ with $c$, $d$, $e$; $2M_S=5$ with $c$) lie below the DMRG extrapolations by approximately $0.001$--$0.32$~mHa. Note that we do not use these small differences as evidence of substantial errors in the published DMRG references. For the $d$ and $e$ active spaces the DMRG energy exceeds CCSD(T) by $0.031020$ and $0.021386$~Ha, and its $E_{\mathrm{var}}$ and $M\to\infty$ entries coincide. The underlying $M$ series shows that this coincidence is not an artifact but true saturation: the energy is unchanged to five decimal places between $M=2048$ and $M=4096$, so the polynomial extrapolation returns the largest-$M$ value. The saturation of the DMRG energies above the corresponding CCSD(T) values motivates further investigation of DMRG convergence in these active spaces. The CCSD(T) entry for $2M_S=5$ in active space $x$ deviates by approximately $0.44$~Ha from the trend in neighboring active spaces and is therefore excluded from quantitative comparisons. The $2M_S=1$ sector is far more strongly multireference and does not fall below the reference for the $c$--$f$ spaces. These remaining differences motivate GPU-accelerated expansion of the determinant space and tests of orbital optimization~\cite{zhang2026absorbingmanybodycorrelationscoreoptimized}. All TrimCI calculations used Hamiltonian or FCIDUMP files obtained from the repositories associated with~\cite{Goings_2022,doi:10.1021/acs.jctc.7b00270}. The raw $2M_S=3$ tabulation reports a single identical $M$ series for all six
active spaces, so no per-space finite-$M$ DMRG energy is available for that sector; the $E_{\mathrm{var}}$ entries are therefore left blank (--) and $\Delta E$ is computed against the $M\to\infty$ values. 
}
\label{tab:trimci_best}
\end{table*}

\clearpage
\newsavebox{\allinonebox} 
\begin{table}[p]
  \centering
  \sbox{\allinonebox}{%
    \scriptsize
    \setlength{\tabcolsep}{1.25pt}%
    \renewcommand{\arraystretch}{1.80}%
    \begin{tabular}{ l l r r l | r r r r | r r r r }
\toprule
& & & & &
\multicolumn{4}{c|}{\textbf{Ground State Energy Estimation}} &
\multicolumn{4}{c}{\textbf{Resources}} \\
\cmidrule(lr){6-9} \cmidrule(lr){10-13}
\makecell{Mole-\\cules} &
\makecell{[Ne,\\No]} &
\makecell{FCI\\size} &
\makecell{Ndets\\(t-sym)} &
\makecell{$\epsilon_{\mathrm{var}}$} &
\makecell{$E_{\mathrm{var}}$\\(Ha)} &
\makecell{$|E_{\mathrm{PT2}}|$\\(Ha)} &
\makecell{$E_{\mathrm{tot}}$\\(Ha)} &
\makecell{$E_{\mathrm{extrap.}}$\\(Ha)} &
\makecell{Wall$^{a}$\\(s)} &
\makecell{CPU$^{a}$\\(s)} &
\makecell{Mem$^{b}$\\(GiB)} &
\makecell{Nodes $\times$\\cores/node$^{b}$} \\
\midrule
\multicolumn{13}{l}{\textbf{P450} \quad $2M_S=1$} \\
\midrule
& [17e, 15o] & 3.22072E+07 & 52{,}477 & 1.00E-05 & -2756.5570738823 & 1.765E-06 & -2756.5570756473 & -2756.5570765083 & 46 & 48{,}097 & 0 & 12 $\times$ 128 \\
& [25e, 23o] & 1.54687E+12 & 665{,}137 & 1.00E-05 & -2756.5856385946 & 4.990E-05 & -2756.5856884927 & -2756.5856895630 & 327 & 143{,}935 & 5 & 1 $\times$ 128 \\
& [33e, 31o] & 7.96980E+16 & 3{,}857{,}227 & 1.00E-05 & -2756.6205930679 & 3.813E-04 & -2756.6209743520 & -2756.6240649212 & 893 & 294{,}658 & 32 & 1 $\times$ 128 \\
& [45e, 41o] & 4.94494E+22 & 20{,}272{,}150 & 1.00E-05 & -2756.7684622377 & 4.677E-03 & -2756.7731390991 & -2756.7761737202 & 2{,}521 & 2{,}055{,}782 & 344 & 8 $\times$ 128 \\
& [47e, 43o] & 7.68907E+23 & 21{,}732{,}956 & 1.00E-05 & -2756.7838936783 & 4.991E-03 & -2756.7888846052 & -2756.7922014899 & 2{,}874 & 2{,}388{,}539 & 361 & 8 $\times$ 128 \\
& [63e, 58o] & 5.81497E+32 & 72{,}612{,}690 & 1.00E-05 & -2756.9840935866 & 1.413E-02 & -2756.9982280291 & -2757.0076668034 & 8{,}067 & 7{,}936{,}514 & 2{,}313 & 32 $\times$ 128 \\
\multicolumn{13}{l}{\textbf{P450} \quad $2M_S=3$} \\
\midrule
& [17e, 15o] & 1.93243E+07 & 32{,}969 & 1.00E-05 & -2756.5569988714 & 8.078E-07 & -2756.5569996791 & -2756.5580885316 & 62 & 22{,}628 & 0 & 1 $\times$ 128 \\
& [25e, 23o] & 1.10490E+12 & 1{,}037{,}289 & 1.00E-05 & -2756.5855764595 & 4.153E-05 & -2756.5856179876 & -2756.5826947264 & 237 & 114{,}506 & 8 & 1 $\times$ 128 \\
& [33e, 31o] & 6.19873E+16 & 2{,}176{,}813 & 1.00E-05 & -2756.6206919020 & 2.726E-04 & -2756.6209645415 & -2756.6209970484 & 331 & 157{,}686 & 16 & 1 $\times$ 128 \\
& [45e, 41o] & 4.07958E+22 & 18{,}661{,}484 & 1.00E-05 & -2756.7817009710 & 4.732E-03 & -2756.7864328282 & -2756.7889815596 & 2{,}031 & 1{,}701{,}174 & 338 & 8 $\times$ 128 \\
& [47e, 43o] & 6.40024E+23 & 20{,}592{,}909 & 1.00E-05 & -2756.7968456136 & 5.268E-03 & -2756.8021132439 & -2756.8051434288 & 2{,}285 & 1{,}886{,}480 & 365 & 8 $\times$ 128 \\
& [63e, 58o] & 5.07237E+32 & 66{,}593{,}162 & 1.00E-05 & -2756.9990437513 & 1.363E-02 & -2757.0126703026 & -2757.0210329019 & 10{,}059 & 8{,}214{,}936 & 1{,}179 & 8 $\times$ 128 \\
\multicolumn{13}{l}{\textbf{P450} \quad $2M_S=5$} \\
\midrule
& [17e, 15o] & 6.83183E+06 & 11{,}331 & 1.00E-05 & -2756.5113608912 & 4.026E-07 & -2756.5113612937 & -2756.5113612991 & 34 & 18{,}611 & 0 & 12 $\times$ 128 \\
& [25e, 23o] & 5.60952E+11 & 359{,}164 & 1.00E-05 & -2756.5071934636 & 2.674E-05 & -2756.5072202082 & -2756.5076865743 & 83 & 107{,}724 & 3 & 1 $\times$ 128 \\
& [33e, 31o] & 3.74227E+16 & 2{,}235{,}219 & 1.00E-05 & -2756.5648166597 & 2.867E-04 & -2756.5651033193 & -2756.5651311598 & 403 & 171{,}551 & 23 & 1 $\times$ 128 \\
& [45e, 41o] & 2.77411E+22 & 17{,}678{,}097 & 1.00E-05 & -2756.7137995735 & 4.510E-03 & -2756.7183095901 & -2756.7198481213 & 2{,}349 & 2{,}123{,}324 & 349 & 8 $\times$ 128 \\
& [47e, 43o] & 4.43093E+23 & 19{,}359{,}154 & 1.00E-05 & -2756.7297313731 & 4.578E-03 & -2756.7343096135 & -2756.7370759224 & 2{,}368 & 1{,}930{,}924 & 366 & 8 $\times$ 128 \\
& [63e, 58o] & 3.85829E+32 & 62{,}610{,}882 & 1.00E-05 & -2756.9271956088 & 1.256E-02 & -2756.9397555783 & -2756.9482330451 & 10{,}377 & 8{,}421{,}981 & 1{,}113 & 8 $\times$ 128 \\
\midrule
\textbf{FeMoco} & [113e, 76o] & 3.64178E+35 & 75{,}554{,}084 & 1.50E-05 & -22139.9794584481 & 5.129E-02 & -22140.0307481988 & -22140.2363692859 & 14{,}653 & 51{,}960{,}944 & 83 & 100 $\times$ 128 \\
\midrule
\makecell{Ru(I)} & [48e, 52o] & 1.81804E+29 & 10{,}073{,}681 & 5.00E-06 & -7361.4638168802 & 7.416E-04 & -7361.4645584556 & -7361.4646502646 & 748 & 515{,}638 & 191 & 24 $\times$ 128 \\
\makecell{Ru\\(XVIII)} & [64e, 56o] (defCD) & 1.89663E+31 & 7{,}339{,}245 & 5.00E-06 & -7475.4399530760 & 4.351E-04 & -7475.4403882065 & -7475.4404335929 & 752 & 524{,}815 & 235 & 24 $\times$ 128 \\
& [100e, 100o] (defCD) & 1.01791E+58 & 10{,}365{,}091$^{t}$ & 5.00E-06 & -7475.5440812063 & 1.395E-03 & -7475.5454766895 & -7475.5457651775 & 6{,}620 & 2{,}582{,}250 & 684 & 24 $\times$ 128 \\
& [150e, 150o] (defCD) & 8.61668E+87 & 16{,}124{,}840$^{t}$ & 1.00E-05 & -7475.6405405986 & 1.182E-02 & -7475.6523632715 & -7475.6565454772 & 38{,}353 & 16{,}747{,}365 & 150 & 8 $\times$ 128 \\
& [64e, 56o] (highCD) & 1.89663E+31 & 7{,}101{,}936 & 5.00E-06 & -7475.4424484746 & 4.098E-04 & -7475.4428583175 & -7475.4429007341 & 383 & 466{,}037 & 193 & 24 $\times$ 128 \\
\midrule
\textbf{Fe2S2} & [30e, 20o] & 2.40374E+08 & 4{,}208{,}596$^{t}$ & 5.00E-06 & -116.6054426024 & 1.222E-04 & -116.6055648123 & -116.6056394161 & 886 & 1{,}957{,}935 & 83 & 28 $\times$ 128 \\
\textbf{Fe4S4} & [54e, 36o] & 8.86296E+15 & 75{,}238{,}321$^{t}$ & 7.00E-06 & -327.0170585057 & 2.895E-02 & -327.0460065195 & -327.1337750658 & 25{,}130 & 70{,}610{,}504 & 3{,}693 & 36 $\times$ 128 \\
\bottomrule
\end{tabular}%
  }%
  \begin{adjustbox}{angle=90, width=\textwidth, height=\textheight}
    \begin{minipage}{\wd\allinonebox}
      \captionsetup{justification=raggedright,singlelinecheck=false}
      \captionof{table}{Summary of SHCI results and computational resources for benchmark systems. Since $\epsilon_{\mathrm{var}}$ was not pushed to our resource limit here, the energies may decrease further and should be read in conjunction with the reported $\epsilon_{\mathrm{var}}$ and $N_{\mathrm{det}}$ values. Because TrimCI requires far less memory than SHCI, it can reach smaller $\epsilon_{\mathrm{var}}$ and thus lower variational energies. $^{a}$Wall and CPU times are cumulative over the orbital-optimization run and the final variational + PT2 run, which may use different numbers of nodes. $^{b}$Peak memory and number of nodes used in the final SHCI run.}
      \label{tab:allinone}
      \par\vspace{6pt}
      \usebox{\allinonebox}
    \end{minipage}
  \end{adjustbox}
\end{table}

\newsavebox{\allinoneTrimbox}
\begin{table}[p]
  \centering
  \sbox{\allinoneTrimbox}{%
    \scriptsize
    \setlength{\tabcolsep}{1.25pt}%
    \renewcommand{\arraystretch}{1.80}%
    \begin{tabular}{ l l r r l | r r r r | r r r r }
\toprule
& & & & &
\multicolumn{4}{c|}{\textbf{Ground State Energy Estimation}} &
\multicolumn{4}{c}{\textbf{Resources}} \\
\cmidrule(lr){6-9} \cmidrule(lr){10-13}
\makecell{Mole-\\cules} &
\makecell{[Ne,\\No]} &
\makecell{FCI\\size} &
\makecell{Ndets\\(t-sym$^{t}$)} &
\makecell{$\epsilon_{\mathrm{var}}$} &
\makecell{$E_{\mathrm{var}}$\\(Ha)} &
\makecell{$E_{\mathrm{PT2}}$\\(Ha)} &
\makecell{$E_{\mathrm{tot}}$\\(Ha)} &
\makecell{$E_{\mathrm{extrap.}}$\\(Ha)} &
\makecell{Wall\\(s)} &
\makecell{CPU\\(s)} &
\makecell{Mem\\(GiB)} &
\makecell{Nodes $\times$\\cores/node} \\
\midrule
\multicolumn{13}{l}{\textbf{P450} \quad $2M_S=1$} \\
\midrule
 & [25e, 23o] & 1.54687E+12 & 290{,}576 & 2.00E-05 & -2756.58559099 & -0.00009060 & -2756.58568159 & -2756.58568399 & 324 & 190{,}723 & 2 & 1 $\times$ 120 \\
 & [33e, 31o] & 7.96980E+16 & 1{,}905{,}663 & 2.00E-05 & -2756.61973616 & -0.00119270 & -2756.62092886 & -2756.62110885 & 1{,}023 & 35{,}829 & 21 & 1 $\times$ 120 \\
 & [45e, 41o] & 4.94494E+22 & 19{,}788{,}193 & 1.00E-05 & -2756.78200991 & -0.00422408 & -2756.78623398 & -2756.78916339 & 2{,}710 & 112{,}665 & 40 & 1 $\times$ 120 \\
 & [47e, 43o] & 7.68907E+23 & 21{,}330{,}547 & 1.00E-05 & -2756.79751615 & -0.00455336 & -2756.80206950 & -2756.80527432 & 2{,}536 & 131{,}562 & 46 & 1 $\times$ 120 \\
 & [63e, 58o] & 5.81497E+32 & 70{,}808{,}170 & 1.00E-05 & -2757.00082950 & -0.01154055 & -2757.01237005 & -2757.01957431 & 5{,}731 & 417{,}440 & 0 & 80 $\times$ 128 \\
\multicolumn{13}{l}{\textbf{P450} \quad $2M_S=3$} \\
\midrule
 & [17e, 15o] & 1.93243E+07 & 13{,}810 & 2.00E-05 & -2756.55699761 & -0.00000206 & -2756.55699967 & -2756.55699963 & 307 & 20{,}174 & 0 & 1 $\times$ 120 \\
 & [25e, 23o] & 1.10490E+12 & 240{,}736 & 2.00E-05 & -2756.58553856 & -0.00007880 & -2756.58561736 & -2756.58561925 & 329 & 10{,}057 & 2 & 1 $\times$ 120 \\
 & [33e, 31o] & 6.19873E+16 & 1{,}605{,}964 & 2.00E-05 & -2756.61994164 & -0.00093525 & -2756.62087689 & -2756.62104519 & 966 & 33{,}821 & 16 & 1 $\times$ 120 \\
 & [45e, 41o] & 4.07958E+22 & 19{,}718{,}410 & 1.00E-05 & -2756.78199906 & -0.00420693 & -2756.78620599 & -2756.78910886 & 2{,}181 & 113{,}679 & 43 & 1 $\times$ 120 \\
 & [47e, 43o] & 6.40024E+23 & 21{,}256{,}360 & 1.00E-05 & -2756.79750480 & -0.00453358 & -2756.80203838 & -2756.80525307 & 2{,}326 & 95{,}135 & 46 & 1 $\times$ 120 \\
 & [63e, 58o] & 5.07237E+32 & 58{,}410{,}081 & 1.00E-05 & -2757.00297936 & -0.01050089 & -2757.01348025 & -2757.02032550 & 5{,}544 & 235{,}602 & 103 & 3 $\times$ 64 \\
\multicolumn{13}{l}{\textbf{P450} \quad $2M_S=5$} \\
\midrule
 & [17e, 15o] & 6.83183E+06 & 7{,}060 & 2.00E-05 & -2756.51136026 & -0.00000107 & -2756.51136133 & -2756.51136132 & 201 & 7{,}762 & 0 & 1 $\times$ 120 \\
 & [25e, 23o] & 5.60952E+11 & 168{,}051 & 2.00E-05 & -2756.53900001 & -0.00005255 & -2756.53905256 & -2756.53905181 & 384 & 16{,}756 & 1 & 1 $\times$ 120 \\
 & [33e, 31o] & 3.74227E+16 & 827{,}868 & 2.00E-05 & -2756.56464811 & -0.00044228 & -2756.56509039 & -2756.56512101 & 761 & 16{,}428 & 6 & 1 $\times$ 120 \\
 & [45e, 41o] & 2.77411E+22 & 13{,}971{,}602 & 1.00E-05 & -2756.71585667 & -0.00291595 & -2756.71877261 & -2756.72076290 & 2{,}317 & 66{,}982 & 30 & 1 $\times$ 120 \\
 & [47e, 43o] & 4.43093E+23 & 15{,}127{,}051 & 1.00E-05 & -2756.73151734 & -0.00315684 & -2756.73467418 & -2756.73688408 & 2{,}325 & 53{,}079 & 31 & 1 $\times$ 120 \\
 & [63e, 58o] & 3.85829E+32 & 43{,}501{,}555 & 1.00E-05 & -2756.93423849 & -0.00786363 & -2756.94210212 & -2756.94761336 & 7{,}677 & 207{,}683 & 84 & 3 $\times$ 64 \\
\midrule
\textbf{FeMoco} & [113e, 76o] & 3.64178E+35 & 116{,}364{,}015 & 8.00E-06 & -22140.26861924 & -0.02728728 & -22140.29590652 & -22140.38676535 & 12{,}590 & 39{,}558{,}432 & 91 & 100 $\times$ 128 \\
\midrule
\textbf{Ru(I)} & [48e, 52o] & 1.81804E+29 & 9{,}334{,}371 & 5.00E-06 & -7361.46388338 & -0.00068612 & -7361.46456951 & -7361.46465032 & 14{,}766 & 409{,}091 & 56 & 3 $\times$ 120 \\
\midrule
\textbf{Ru(XVIII)} & [64e, 56o]\,(defCD) & 1.89663E+31 & 7{,}524{,}522 & 5.00E-06 & -7475.43993722 & -0.00044981 & -7475.44038703 & -7475.44043516 & 11{,}707 & 279{,}751 & 144 & 2 $\times$ 120 \\
\textbf{Ru(XVIII)} & [100e, 100o]\,(defCD) & 1.01791E+58 & 10{,}365{,}084$^{t}$ & 5.00E-06 & -7475.54408121 & -0.00139762 & -7475.54547882 & -7475.54576956 & 93{,}395 & 1{,}366{,}534 & 115 & 3 $\times$ 64 \\
\textbf{Ru(XVIII)} & [150e, 150o]\,(defCD) & 8.61668E+87 & 16{,}104{,}910$^{t}$ & 1.00E-05 & -7475.64053447 & -0.01182781 & -7475.65236228 & -7475.65662666 & 89{,}648 & 15{,}128{,}859 & 15 & 100 $\times$ 128 \\
\textbf{Ru(XVIII)} & [64e, 56o]\,(highCD) & 1.89663E+31 & 7{,}462{,}114 & 5.00E-06 & -7475.44241588 & -0.00043926 & -7475.44285514 & -7475.44290215 & 18{,}467 & 323{,}306 & 145 & 2 $\times$ 120 \\
\midrule
\textbf{Fe2S2} & [30e, 20o] & 2.40374E+08 & 8{,}011{,}000$^{t}$ & 2.00E-06 & -116.60556142 & -0.00003653 & -116.60559795 & -116.60561525 & 3{,}692 & 441{,}517 & 2 & 3 $\times$ 64 \\
\midrule
\textbf{Fe4S4} & [54e, 36o] & 8.86296E+15 & 353{,}799{,}788$^{t}$ & 4.00E-06 & -327.04484554 & -0.02262823 & -327.06747376 & -326.49568876$^{\dagger}$ & 6{,}493 & 6{,}870{,}106 & 122 & 300 $\times$ 128 \\
\bottomrule
\end{tabular}%
  }%
  \begin{adjustbox}{angle=90, width=\textwidth, height=\textheight}
    \begin{minipage}{\wd\allinoneTrimbox}
      \captionsetup{justification=raggedright,singlelinecheck=false}
      \captionof{table}{Summary of TrimCI results and computational resources for benchmark systems. For resource accounting, this table reports individual TrimCI calculations that may differ from those yielding the lowest variational energies in Table~\ref{tab:trimci_best}. Since $\epsilon_{\mathrm{var}}$ was not necessarily pushed to the computational resource limit, the reported energies may decrease further and should be interpreted in conjunction with the corresponding $\epsilon_{\mathrm{var}}$ and $N_{\mathrm{det}}$ values. Because TrimCI requires substantially less memory than SHCI, it can reach smaller $\epsilon_{\mathrm{var}}$ and potentially lower variational energies. The dagger ($\dagger$) indicates substantial uncertainty in the extrapolated energy. The last column lists the number of nodes $\times$ cores per node used for each calculation. $^{t}$$N_{\mathrm{det}}$ obtained with time-reversal symmetry.}
      \label{tab:allinoneTrim}
      \par\vspace{6pt}
      \usebox{\allinoneTrimbox}
    \end{minipage}
  \end{adjustbox}
\end{table}

\end{document}